\documentclass[fleqn,usenatbib]{mnras}

\usepackage{newtxtext,newtxmath}
\usepackage[T1]{fontenc}

\DeclareRobustCommand{\VAN}[3]{#2}
\let\VANthebibliography\thebibliography
\def\thebibliography{\DeclareRobustCommand{\VAN}[3]{##3}\VANthebibliography}

\usepackage{graphicx}	% Including figure files
\usepackage{amsmath}	% Advanced maths commands
\usepackage{amssymb}	% Extra maths symbols
\newcommand{\ergcms}{erg s$^{-1}$ cm$^{-2}$}

\title[Globular Cluster ULXs in M87]{X-Ray Spectroscopy of Newly Identified ULXs Associated With M87's Globular Cluster Population}

\author[K.C. Dage et al]{
Kristen C. Dage,$^{1}$\thanks{E-mail: kcdage@msu.edu}
Stephen E. Zepf,$^{1}$, 
Erica Thygesen, $^{1}$
Arash Bahramian, $^{2}$
\newauthor
Arunav Kundu $^3$
Thomas J. Maccarone, $^4$
Mark B. Peacock,$^{1}$ 
Jay Strader$^{1}$ 
\\
$^{1}$Department  of  Physics  and  Astronomy,  Michigan  State  University,  East Lansing, MI 48824\\
$^{2}$ International Centre for Radio Astronomy Research $--$ Curtin University, GPO Box U1987, Perth, WA 6845, Australia\\
$^{3}$Eureka Scientific, Inc., 2452 Delmer Street, Suite 100 Oakland, CA 94602, USA\\
$^{4}$Department of Physics, Box 41051, Science Building, Texas Tech University, Lubbock, TX 79409-1051, USA \\
\\
}

\date{Accepted XXX. Received YYY; in original form ZZZ}

\pubyear{2020}

\begin{document}
\label{firstpage}
\pagerange{\pageref{firstpage}--\pageref{lastpage}}
\maketitle

\begin{abstract}
We have identified seven ultraluminous X-ray sources (ULXs) which are coincident with globular cluster candidates (GCs) associated with M87. ULXs in the old GC environment represent a new population of ULXs, and ones likely to be black holes.  In this study we perform detailed X-ray spectroscopic followup to seven GC ULXs across a wealth of archival \textit{Chandra} observations and long time baseline of 16 years.  This study brings the total known sample of GC ULXs to 17. Two of these sources show variability in their X-ray luminosity of an order of magnitude over many years, and one of these sources shows intra-observational variability on the scale of hours. While the X-ray spectra of the majority of globular cluster ULXs are best fit by single component models, one of the sources studied in this paper is the second GC ULX to be best fit by a two component model.  We compare this new sample of GC ULXs to the previously studied sample, and compare the X-ray and optical properties counterparts across the samples. We find that the clusters that host ULXs in M87 have metallicities from $g-z$=1.01 to  $g-z$=1.70. The best-fit power-law indices of the X-ray spectra range from $\Gamma$=1.37-2.21, and the best fit inner black-body disk temperatures range from kT=0.56-1.90 keV.
\end{abstract}


%%%%%%%%%%%%%%%%%%%%%%%%%%%%%%%%%%%%%%%%%%%%%%%%%%

%%%%%%%%%%%%%%%%% BODY OF PAPER %%%%%%%%%%%%%%%%%%

\section{Introduction}

M87 is an elliptical galaxy in the Virgo group, and home to thousands upon thousands of globular clusters (GCs) \citep{Harris09}. While M87 is best known for hosting a super massive black hole with high energy jets, the galaxy is filled with hot gas with energies up to 2.5keV (e.g. \citealt{1978ApJ...226L.107F}). This hot gas makes studying fainter source populations difficult, but the brightest X-ray sources can be readily found and studied above the background from the hot gas.

The brightest X-ray sources are known as ultraluminous X-ray sources (ULXs) and are sources which typically exceed $\sim 10^{39}$ erg/s (i.e. above the Eddington limit for a 10 $M_\odot$ black hole) \citep{fabbiano06}. These sources are thought to be a $\sim$ stellar mass compact object plus a donor star accreting in a different regime of physics than typical X-ray binaries. 
  
\citet{2009MNRAS.398.1450K} show that there are likely to be at least two classes of ULXs observed  within the young star-forming galaxies. One class is ULXs that are persistently below $3\times10^{39}$ erg/s that show relatively hot thermal spectra where the source variations are consistent with $L\propto T^4$. The other set of sources shows power law dominated spectra, and a quasi-thermal soft excess \citep[e.g.,][]{2018ApJ...856..128W}.

Many ULXs have massive donor stars \citep[e.g.,][]{gladstone13}, and the majority of them have been found associated with the star-forming regions of galaxies \citep[e.g.,][and references therein]{2019arXiv191204431B}. 
Of the ULXs identified in star-forming regions of galaxies, six have had a neutron star (NS) identified as the primary accretor in the system via detection of pulsations, and many more systems like this are thought to also be powered by a NS \citep{song20, jithesh20}, which is perhaps highly magnetized
and/or beamed \citep{king08, brightman18}. 
However, ten sources with ultraluminous X-ray luminosities that were bright on the scale of years have been found to be associated with globular clusters (GCs) in nearby elliptical galaxies,  \citep{maccarone07, irwin10, shih10, maccarone11, roberts12,  dage19a}, as well as two GC sources which flared to above $10^{39}$ erg/s on the timescale of minutes \citep{Irwin16}. As old stellar populations, GCs are not home to the types of massive donor stars typically associated with ULXs, and the ULXs associated with the two types of galaxies are likely distinct populations \citep[e.g.][]{irwin03}. At least one GC ULX is likely to be a black hole (BH) accreting off of a white dwarf (see \citealt{zepf08, steele11,peacock12neb, dage19b} and references therein).

A systematic study of the known GC ULXs by \citealt{dage19a} revealed that sources were either best fit by a power-law model, or by black-body disk model. 
The accretion disk models fits  fell into one of two groups: either the sources were better fit by a higher disk temperature or had a lower disk temperature with the spectral parameter constant while varying in luminosity. There is a likely correlation between the disk temperature  and whether or not the source has optical emission, as the optical spectra of the low temperature sources show forbidden optical emission lines beyond the globular cluster continuum, and a number of the higher temperature sources have no emission beyond the cluster continuum \citep{dage19a}. However, this conclusion is tentative because the number of GC ULXs is still small.

Understanding the number and nature of these rare sources has implications not just for accretion physics, but for addressing the question of black holes in extragalactic globular clusters. While black holes (or good black hole candidates) have begun to be observed in Galactic globular clusters (see \citealt{strader12, 2013ApJ...777...69C, MillerJones15, giesers18, shishkovsky18, giesers19} for recent examples), the methods used to detect quiescent BHs in Galactic GCs cannot be used to search for BHs in extragalactic GCs. Thus, searching for ULXs in extragalactic GCs is the best option for identifying BH candidates in GCs outside our own Galaxy.

One very likely scenario to form at least some black hole-black hole binaries (BBHs) are GCs \citep{abbott16, 2019PhRvD.100d3027R}.  Clusters represent an older population of stars, which will have initially produced many BHs. Because of the dynamic environment, the cross section for interaction is high and binaries can form easily. Once formed, BHs may possibly drive the cluster evolution \citep{giersz19, kremer19}. Observational studies of BHs in GCs have a large impact both on understanding cluster evolution, as well as giving clues to the conditions that BBHs form in.  ULX sources in GCs offer a complementary path to help answer the question of numbers of BHs in GCs, as well as touch on the question of the origins of the progenitors of LIGO sources.

In this body of work, we use  \textit{Chandra} observations of M87 and its surrounding regions to identify ultraluminous X-ray sources coincident with GCs. The observations and analysis are laid out in Section \ref{sec:obs}, major results are highlighted in Section \ref{sec:res}, and the implications of this work are discussed in Section \ref{sec:conc}.

\section{Observations and Analysis}
\label{sec:obs}
M87 has a wealth of archival data available, including 28 \textit{Chandra} ACIS observations with the duration of 15~ks or longer (see Table \ref{chandradata}). M87 is host to almost 15,000 globular clusters, and we utilize the photometric globular cluster catalogs which have conservatively modeled and removed contaminant populations from  \citealt{2016MNRAS.455..820O} as well as catalogs of confirmed globular clusters \citep{strader11, caldwell14}. 

\begin{table*}
\centering
\caption{Chandra observations of M87 with observations greater than 15~ks. All are ACIS-S with no grating unless indicated otherwise.}
\label{chandradata}
\begin{tabular}{llllllll}
\hline \hline
\multicolumn{1}{|l|}{ObsID} & \multicolumn{1}{l|}{Date} & \multicolumn{1}{l|}{Exposure (ks)} &
\multicolumn{1}{l|}{Sources in FOV} &
\multicolumn{1}{l|}{ObsID} & \multicolumn{1}{l|}{Date} & \multicolumn{1}{l|}{Exposure (ks)} &
\multicolumn{1}{l|}{Sources in FOV} \\ \hline
241 (HETG) & 2000-07-17  & 40.0   & 1                        & 15180   (ACIS-I)                    & 2013-08-01                & 140.0     & SC302   \\
352                         & 2000-07-29                & 40.0    & L18, J04, 1-4                       & 16585          (ACIS-I)            & 2014-02-19                & 45.0  & SC302                            \\
2707                        & 2002-07-06                & 105.0     &  L18, J04, 1-4                      & 16586          (ACIS-I)            & 2014-02-20                & 50.0       & SC302                       \\
3717                        & 2002-07-05                & 23.0        & L18, J04, 1-3                   & 16587    (ACIS-I)                  & 2014-02-22                & 38.0          & SC302                    \\
4007                        & 2003-11-21                & 40.0      & None                     & 16590      (ACIS-I)                & 2014-02-27                & 38.0     & SC302                         \\
5826 (ACIS-I)                       & 2005-03-03                & 140.0       & L18, J04, 1-4                      & 16591     (ACIS-I)                 & 2014-02-27                & 24.0       & SC302                       \\
5827 (ACIS-I)                       & 2005-05-05                & 160.0   &L18, J04, 1-4                          & 16592       (ACIS-I)               & 2014-03-01                & 36.0        & SC302                      \\
5828  (ACIS-I)                      & 2005-11-17                & 33.0       & L18, J04, 1-4                       & 16593    (ACIS-I)                  & 2014-03-02                & 38.0         & SC302                     \\
6186    (ACIS-I)                    & 2005-01-31                & 48.0      &L18, J04, 1-3                     & 18781                      & 2016-02-24                & 45.0 & J04, 1-3                             \\
7210  (ACIS-I)                      & 2005-11-16                & 32.0            & L18, J04, 1, 3,4                  & 18782                      & 2016-02-26                & 37.0                  &  J04, 1-3             \\
7211   (ACIS-I)                     & 2005-11-16                & 17.0         & L18,  J04, 1-3                  & 18783                      & 2016-04-20                & 40.0                         &  J04, 2,3    \\
11783                       & 2010-04-13                & 30.0            &L18               & 18836                      & 2016-04-28                & 43.0             &     J04, 2,3            \\
15178 (ACIS-I)                      & 2014-02-17                & 47.0      &SC302                     & 18838                      & 2016-05-28                & 62.0           & J04, 2,3                   \\
15179   (ACIS-I)                    & 2014-02-24                & 44.0                &SC302         & 18856                      & 2016-06-12                & 28.0  & J04, 2,3         \\                      
\hline
\end{tabular}
\end{table*}
\subsection{Identification of Globular Cluster ULXs}

We reprocessed the \textit{Chandra} data using \textsc{ciao} version 4.11 \citep{2006SPIE.6270E..1VF} \texttt{chandra\_repro} and CalDB version 4.10.  We  filtered out any background flares. We used the \texttt{Wavdetect} function within \textsc{Ciao} to identify X-ray point sources  across the observations using 0.5-7.0 keV images. We used an exposure map centered at 2.3 keV, and wavelet scales at 1.0, 2.0, 4.0, 8.0 and 16.0 pixels. %First we ran wavdetect on the initial images to find any background sources to exclude, using an enclosed count fraction (e.c.f.) of 0.9, and significance threshold of $10^{-5}$.
Once the images had been cleaned and filtered ,we ran \texttt{wavdetect} using an enclosed count fraction (e.c.f.) of 0.3 \footnote{We chose a low value because of the high background.} and significance threshold of 10$^{-6}$ which corresponds to about one false detection per chip.
We then used the \texttt{srcflux} function to estimate unabsorbed X-ray fluxes assuming a fixed Galactic line of sight $N_H$ of 2.54 $\times 10^{20}$ cm$^{-2}$ \citep{1990ARA&A..28..215D} and a fixed power-law index of $\Gamma=1.7$, which is the typical power-law index for X-ray binaries. 

\begin{figure}

\includegraphics[width=3.5in]{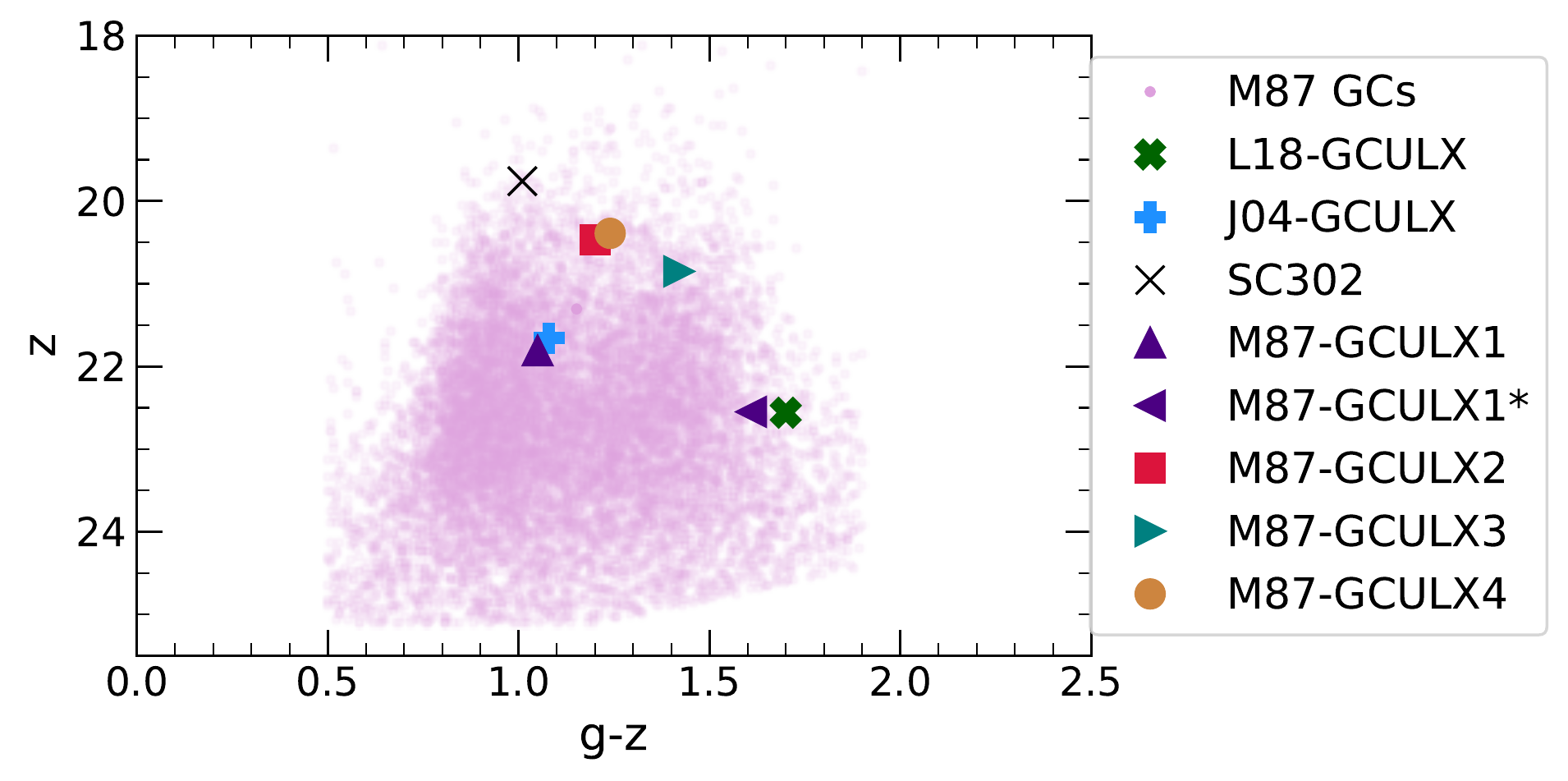}

\caption{$z$ versus $g-z$ for M87 globular cluster photometric candidates from \citealt{jordan09,2016MNRAS.455..820O}, with color and magnitude of GC candidates hosting ULXs plotted overtop.} 
\label{fig:z_gz}
\end{figure}

\begin{figure*}

\includegraphics[width=3.2in]{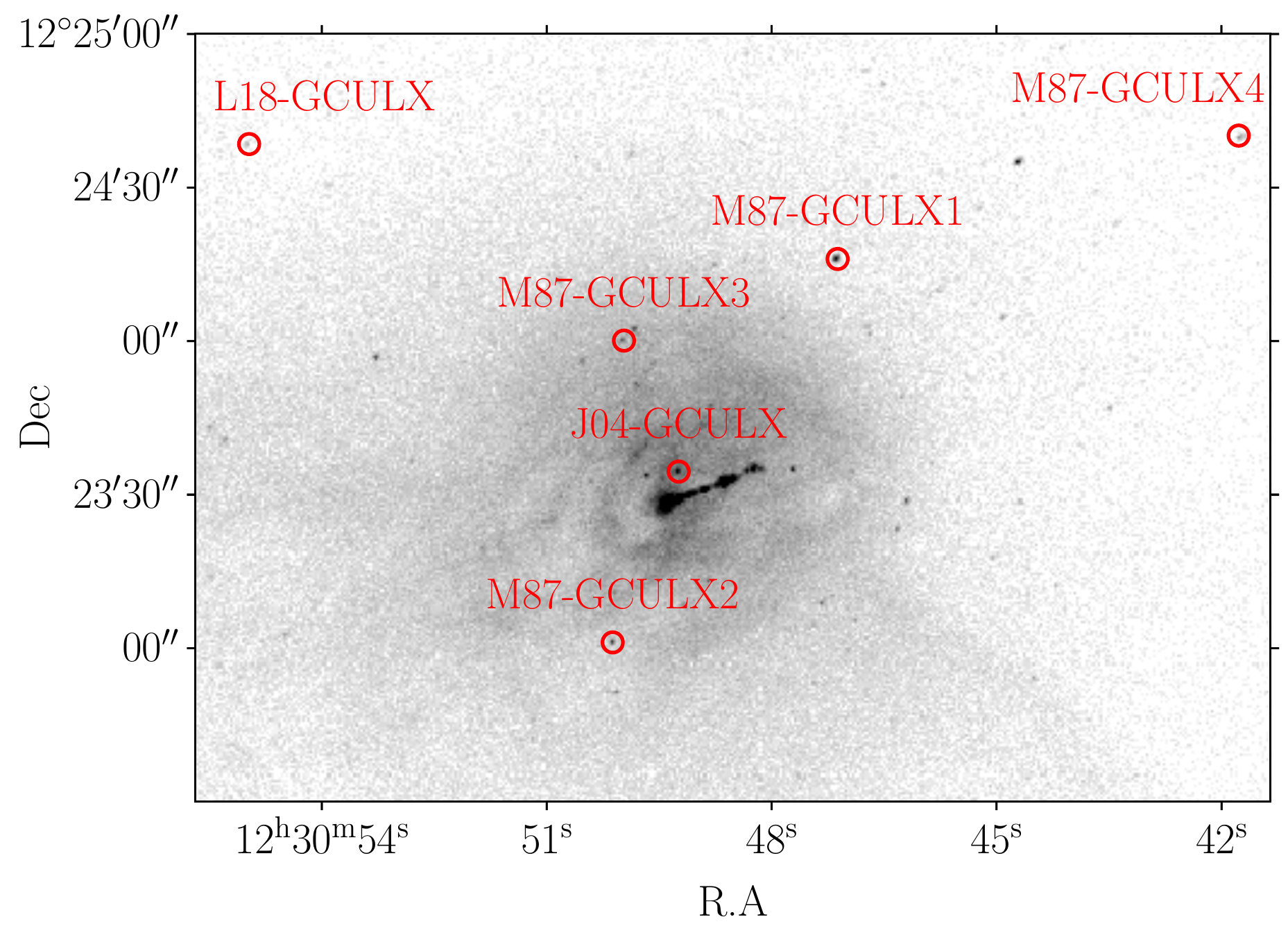}
\includegraphics[width=3.2in]{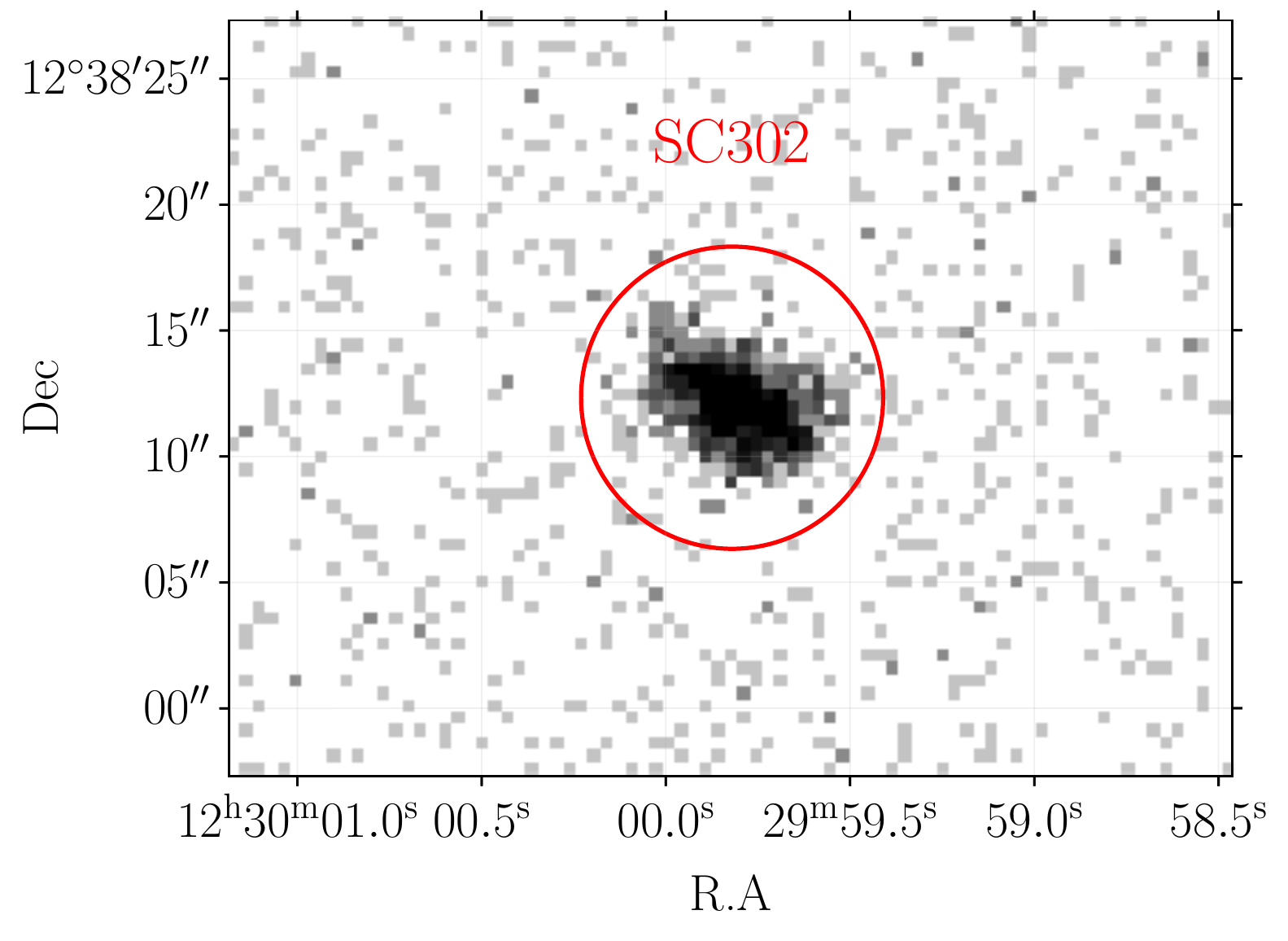}
\caption{\emph{Left} - 0.5-7.0 keV band image of M87 (from ObsID 2707) and four of the ULXs identified and studied in this work. \emph{Right} -  X-ray 0.5-7.0 keV band images of  SC302 (ObsID 15180). This source is observed at high ($>4\arcmin$) off-axis angles, thus shows non-circular point spread functions.}
\label{fig:field1}
\end{figure*}

The X-ray luminosity of the sources was estimated using a distance of 16.8 Mpc \citep{macri}, and selected sources which reached an estimated X-ray luminosity of ~$7\times 10^{38}$ erg/s or above (as using a fixed model tends to underestimate the flux slightly) at one point during the observations. This allowed us to be sensitive to transient ULX sources which may have only brightened once during the sequence of observations, as well as persistent ones.  We used \textsc{topcat} \citep{2005ASPC..347...29T} to match the globular cluster catalogs \citep{jordan09, 2016MNRAS.455..820O} to the ULX sources with a 0\farcs8 tolerance\footnote{0\farcs8 is the 90\% uncertainty radius of \textit{Chandra}'s absolute astrometry: \url{https://cxc.harvard.edu/cal/ASPECT/celmon/}}.
This provided us with a sample of seven ULXs overlapping with globular cluster candidates (Table \ref{sources} and Figure \ref{fig:z_gz}). Several of these sources were also identified in \citealt{2004ApJ...613..279J, Luan18}. One of these sources is located in the Virgo cold front ($\sim$19\arcmin\ from the galaxy center) and another  is located in a field $\sim$8\arcmin\ from the galaxy center. The remaining five sources are near the galaxy center. The closest source to the active jet region is $\sim$ 6\farcs8 away, and the farthest source $\sim$ 130\arcsec\ from the centre (see Figure \ref{fig:field1}). A comparison of the host clusters colors and z-magnitudes (de-reddened \footnote{\url{https://irsa.ipac.caltech.edu/cgi-bin/bgTools/nph-bgExec}}) to other clusters in M87  are plotted in Figure \ref{fig:z_gz}.%Given the high energy levels of the dust, this made it difficult to follow up on the sources closest to the jet in X-rays, however, we perform detailed spectroscopic follow up on the other two new sources.

\begin{table*}
\caption{RA and Dec of identified GC ULXs within M87's system. Photometric values (de-reddened z-band magnitude and g-z color) from \citealt{jordan09,2016MNRAS.455..820O}, Strader et al in prep. Sources marked with $^*$ are found in Strader et al in prep., $^o$ from \citealt{2016MNRAS.455..820O}, and $\dagger$ were identified in \citep{jordan09}. $r_h$ values are from \citealt{jordan09} or measured from HST data using lshape. M87-GCULX1 was matched to two clusters and we present values for both cluster parameters \citep{larsen99}. }
\label{sources}
\begin{tabular}{ccccccc}
 \hline \hline
 Name      & RA           & Dec          & z-band     & g-z       & $r_h$ & $L_X$ \\ 
           &              &              & (AB mag) & (AB mag)  & (pc)  & ($\times 10^{39}$ erg/s)\\
 \hline
CXOU J122959.82+123812.33 (SC302) $^*$ & 12:29:59.82 & +12:38:12.33 & 19.76 &1.01 & - &1.65 \\
CXOU J123054.97+122438.51 (L18-GCULX)  $\dagger$ & 12:30:54.97 & +12:24:38.51 & 22.56 & 1.70 & 1.88& 1.05\\
CXOU J123049.24+122334.52 (J04-GCULX)  $\dagger$& 12:30:49.24 & +12:23:34.52& 21.66 & 1.08 & 2.20& 0.80\\   
CXOU J123047.12+122416.06 (M87-GCULX1) $^o$ $\dagger$ & 12:30:47.12 & +12:24:16.06 & 21.73/22.55 & 1.05/1.61 &3.69/1.96 &1.64\\
CXOU J123050.12+122301.14 (M87-GCULX2) $^o$ $\dagger$ & 12:30:50.12 & +12:23:01.14 & 20.47&1.29& 1.82&1.01\\ 
CXOU J123049.97+122400.11 (M87-GCULX3)  $^o$ $\dagger$ & 12:30:49.97 & +12:24:00.11 & 20.85& 1.42 & 2.53&1.14 \\
CXOU J123041.77+122440.16 (M87-GCULX4) $^o$ & 12:30:41.77 & +12:24:40.16 & 20.41& 1.24 & 2.70&1.08\\  \hline 
%CXOUJ123017.77+122545.15 (M87-GCULX5) $^o$& 12:30:17.77 & +12:25:45.15 & 21.78& 1.58 & 1.74\\ \hline
\end{tabular}
\end{table*}

%\begin{figure*}
%  
%        \includegraphics[width=3.5in]{SC302.pdf} % %first figure 
%    \caption{X-ray 0.5-7.0 keV band images of  SC302 ( %ObsID 15180) ). This sources is observed at high %($>4\arcmin$) off-axis angles, thus shows non-circular %point spread functions.}
%    \label{fig:field2}
%\end{figure*}

\subsection{X-ray Spectroscopy}
Once the seven GC ULXs were identified, we extracted the source spectra in every observation that had a point source associated with the source's position using \textsc{ciao} with 1.2\farcs circular source extraction regions, with a series of similar sized region files placed around the source to extract background spectra, taking care to avoid any neighbouring X-ray sources.

After extracting all spectra, we used \textsc{combine\_spectra} to combine the spectra, and binned by 20 using \textsc{dmgroup} to have one deep spectrum for every source. Previous studies of GC ULXs have established that they are fit well either by an absorbed multi-colored disk \citep[\texttt{tbabs*diskbb} in Xspec; e.g.,][]{1984PASJ...36..741M} and/or an absorbed power-law \citep[\texttt{tbabs*pegpwrlw} in Xspec; e.g.,][]{irwin10,shih10, maccarone11,roberts12}, or an additive combination of the two models \citep[\texttt{tbabs*(diskbb+pegpwrlw)} in Xspec; e.g.,][]{maccarone07}.
%with one source being best fit by an additive combination of the two models \texttt{tbabs*(diskbb+pegpwrlw)} \citep{maccarone07}. 
We used these models to fit the combined spectra of the sources in our sample of GC ULXs in M87 and used F-Test \footnote{\url{https://heasarc.gsfc.nasa.gov/xanadu/xspec/manual/node83.html}} to determine if the two-component model was likely to be a better fit. Fit results and statistics (for combined spectra) are presented in Table \ref{deepfits}. We also considered effects of a secondary absorption column in each source. However we found that in all cases this second column is consistent with zero and thus used a single absorption column.

%We fit those models to the deep spectra of the M87 sources and used F-Test \footnote{\url{https://heasarc.gsfc.nasa.gov/xanadu/xspec/manual/node83.html}} to determine if the two-component model was truly a better fit.  These values are presented in Table \ref{deepfits}.
\begin{table*}
\caption{Single component model fits to the stacked spectra of M87's GC ULXs, with the F-Test comparison to the two component models for the best single-component model tabulated below (lower F-test probability indicates a higher likelihood that adding a second component improves the fit).  $\Gamma_\text{PL}$ represent power-law photon index, kT$_{\text{in}}$ is the inner disk temperature in the disk blackbody model, $\chi_\nu^2$ indicates reduced $\chi^2$, and d.o.f. stands for degrees of freedom in the fit. }
\label{deepfits}

\begin{tabular}{lccccccc}  \hline \hline
            & \multicolumn{3}{c}{power-law}                                &&  \multicolumn{3}{c}{diskbb}   \\    
\cline{2-4}\cline{6-8}\\
Source      & $\Gamma_\text{PL}$ & $\chi_\nu^2$/d.o.f. & F-Test prob. (\%) && kT$_{\text{in}}$ (keV)   & $\chi_\nu^2$/d.o.f.  & F-Test prob. (\%) \\ 
\hline 
SC302       & 2.07 ($\pm$ 0.05)  & 2.00/110            & --                         && 0.92 ($\pm$ 0.04)        & 0.98/110             & 0.8   \\
L18-GCULX   & 2.03 ($\pm$ 0.11)  & 1.11/96             & --                         && 0.81 ($\pm 0.09$)        & 1.03/96              & 1.5  \\
J04-GCULX   & 2.21 ($\pm$ 0.09)  & 1.06/147            & 1.4                        && 0.74 ($\pm$ 0.07)        & 1.26/147             & --  \\
M87-GCULX1  & 2.17 ($\pm$ 0.05)  & 1.35/120            & $\ll0.01$                   && 0.67 ($\pm$ 0.04)        & 1.39/120             & --  \\
M87-GCULX2  & 1.52 ($\pm$ 0.11)   & 1.00/106            & --                         && 1.44 ($\pm$ 0.21)        & 0.95/106             & 20.6 \\
M87-GCULX3  & 1.36 ($\pm$ 0.09)  & 0.91/117            & 7.7                        && 1.90 ($\pm$ 0.28)        & 0.94/117             & --  \\
M87-GCULX4  &  1.37 ($\pm 0.15$) & 1.02/44             & 23.8                       && 1.86 ($^{+0.56}_{-0.38}$)& 1.36/44              & -- \\ 
\hline    
\end{tabular}
\end{table*}

%\begin{tabular}{lllllll}  \hline \hline
%Source   & Gamma              & $\chi ^2$/d.o.f.    & F-Test (power-law)   & kT    (keV)             & %$\chi ^2$/d.o.f.    & F-Test (diskbb) \\ \hline 
%SC302  & 2.07 ( $\pm$ 0.05) & 2.00/110 & 66.99    & 0.92 ($\pm$ 0.04) & 0.98/110 & 5.03   \\
%L18-GCULX & 2.03 ($\pm$ 0.11)  & 1.11/96  & 8.28   & 0.81 ($\pm 0.09$) & 1.03/96  & 4.36  \\
%J04-GCULX  & 2.21 ($\pm$ 0.09)  & 1.06/147 & 4.40& 0.74 ($\pm$ 0.07) & 1.26/147 & 19.03   \\
%M87-GCULX1& 2.17 ($\pm$ 0.05)  & 1.35/120 & 36.32 & 0.67 ($\pm$ 0.04) & 1.39/120 & 39.21  \\
%M87-GCULX2& 1.52($\pm$ 0.11)   & 1.00/106 & 1.60  & 1.44 ($\pm$ 0.21) & 0.95/106 & 4.23   \\
%M87-GCULX3& 1.36 ($\pm$ 0.09)  & 0.91/117 & 4.48   & 1.90 ($\pm$ 0.28) & 0.94/117 & 2.62  \\
%M87-GCULX4  &  1.37 ($\pm 0.15$)   &     1.02/44         &    1.48      &    1.86 %($^{+0.56}_{-0.38}$)& 1.36/44  & 8.67\\ \hline    
%\end{tabular}
%\end{table*}

%We found that none of the sources needed a second absorption component to fit the data. The best fit value for the second absorption component was consistent with zero. 

We also performed spectral analysis on the individual observations for each source, modeling extracted spectra with \texttt{tbabs*diskbb} and \texttt{tbabs*pegpwrlw}. Any source spectrum with $\geq100$ counts was binned by 20 events per bin.  For source spectra with less than 100 counts, we binned them by 1 count per bin and fit the same models using W-statistics \citep{1979ApJ...228..939C}. J04-GCULX, M87-GCULX2 and M87-GCULX3 were especially challenging to fit in X-ray due to their proximity to the center of the jet, and the high background. All individual spectra of M87GCULX3 were entirely fit with W-statistics. 

The results of our spectral analysis and the best-fit values for each of the sources are presented in Tables \ref{sc302fits}-\ref{ulxfits-GCULX4}. %We also plot the spectral fits from the stacked observations for each of the two models for every source (see Appendix \ref{spectra}) except M87-GCULX3, which was fit with W-stats.

Among the two sources farther away from the galaxy, SC302 was well monitored over a short period of time in the Virgo Cold Front (ObsIDs 15178-16593). 

M87-GCULX4 was not in any observation after ObsID 7210 because it was off the chip for the remainder of the observations. We used \textsc{pimms} to estimate upper limits for  M87-GCULX3 in ObsID 352, and for M87-GCULX2 in ObsID 18783, using the value of the best fit power-law index from the longest observation. The upper limit flux estimates based on the count rates are included in  Table \ref{uplims}, as well as Tables \ref{ulxfits-SC302}-\ref{ulxfits-GCULX4}.

\begin{table*}
\centering
    \caption{\textit{Chandra} Fit Parameters and Fluxes (0.5-8 keV) for spectral best fit single-component models,  \texttt{tbabs*pegpwrlw} and\texttt{tbabs*diskbb} for SC302. Hydrogen column density ($N_H$) frozen to 2.54 $\times 10^{20}$cm$^{-2}$. All fluxes shown are unabsorbed in the 0.5-8keV band. }
\label{ulxfits-SC302}
\begin{tabular}{ccccccccc}
\hline
\hline
\multicolumn{9}{c}{SC302}\\ 
\hline
                    & \multicolumn{3}{c}{\textsc{tbabs*pegpwrlw}} && \multicolumn{4}{c}{\textsc{tbabs*diskbb}} \\ 
\cline{2-4}\cline{6-9}\\
ObsID (Date)       & $\Gamma$           &   $\chi^2_{\nu}$/d.o.f. & PL Flux && $T_{in}$ &  Disk Norm &  $\chi^2_{\nu}$/d.o.f. &  Disk Flux  \\
                   &                    &               &($10^{-14}$ \ergcms)&& (keV)&($10^{-3}$)&&($10^{-14}$ \ergcms)\\ 
\hline
15178 (2014-02-17) & 2.1 ($\pm$0.3)  & 1.32/7         & 6.6 ($\pm$ 0.9) && 0.9 $(^{+0.2}_{-0.1})$ & 4.5 ($^{+4.7}_{-2.3}$ &1.19/7 &  5.0 ($\pm$0.7)\\ 
15179 (2014-02-24) & 2.1 ($\pm$ 0.3) & 2.25/9         & 9.1 ($\pm$1.1)  && 0.8 ($\pm$0.1) & 8.7($^{+6.9}_{-3.9}$ &1.35/9& 7.0 ($\pm$0.7) \\
15180 (2013-08-01) & 2.0 ($\pm$0.1)  & 1.57/53        & 11.3 ($\pm$0.1) && 1.0($\pm$0.1)&4.9 ($\pm$1.1)&0.83/53&9.9 ($\pm$0.6)\\
16585 (2014-02-19) & 2.1 ($\pm$0.3)  & 1.19/6         & 6.8 ($\pm$1.0)  && 0.8 ($\pm$0.2)& 8.3 ($^{+8.9}_{-6.5}$&0.29/6&4.9($\pm0.7$))\\
16586 (2014-02-20) & 1.9 ($\pm$0.5)  & 0.57/6         & 7.4 ($\pm$1.5)  && 0.8 ($^{+0.3}_{-0.2}$)&$\leq$6.8 & 0.73/6&4.9 ($^{+1.1}_{-8.0}$) \\ 
16587 (2014-02-22) & 2.1 $(\pm 0.4)$ & 1.64/6         & 7.3 ($\pm$ 1.1) && 0.9 ($^{+0.3}_{-0.2}$) &4.5 ($^{+6.5}_{-2.7}$) & 1.92/6 & 5.4 ($\pm$0.9) \\
16590 (2014-02-27) & 1.9 ($\pm 0.3$) & 1.23/5         & 6.6 ($\pm$1.0)  && 1.0 ($^{+0.3}_{-0.2}$)&2.4 ($^{+3.2}_{-1.5}$) &1.10/5&5.1 ($\pm$0.9)   \\
16591 (2014-02-27) & 2.1 ($\pm$ 0.9)  & 0.22/2         & 8.4 ($^{+3.9}_{-1.7}$) && 0.7 ($^{+0.8}_{-.25}$)&$\leq$ 11.9 & 0.55/2 & 5.5 ($^{+2.9}_{-1.1}$)  \\ 
16592 (2014-03-01) & 1.7 ($\pm$0.5)  & 0.42/4         & 7.9 ($^{+2.0}_{-1.5}$) && 1.0 ($^{+0.5}_{-0.3}$)&$\leq$2.6 & 0.23/4 & 5.5 ($^{+1.6}_{-1.1}$)\\
16593 (2014-03-02) & 1.9 ($^{+0.5}_{-0.4}$) & 0.63/4 & 6.6 ($^{+1.4}_{-1.1}$) && 1.0 ($^{+0.4}_{-0.3}$)& $\leq$2.9 & 1.03/4 & 4.7 ($^{+1.2}_{-0.5}$)\\ \hline
\end{tabular}
\label{sc302fits}
\end{table*}

\begin{table*}
\centering
    \caption{\textit{Chandra} Fit Parameters and Fluxes (0.5-8 keV) for spectral best fit single-component models,  \texttt{tbabs*pegpwrlw} and\texttt{tbabs*diskbb} for L18-GCULX. Hydrogen column density ($N_H$) frozen to 2.54 $\times 10^{20}$cm$^{-2}$. Lower count observations fit with W-stat have their statistics presented in parentheses. All fluxes shown are unabsorbed. Values marked with $\dagger$ are upper-limits based on the count rate of a non-detection (see Table \ref{uplims}), values marked with - were detections but could not be fit with the \texttt{diskbb} model.}
\label{ulxfit-l18gculX}
\begin{tabular}{ccccccccc}
\hline
\hline
\multicolumn{9}{c}{L18-GCULX}\\ 
\hline
                    & \multicolumn{3}{c}{\textsc{tbabs*pegpwrlw}} && \multicolumn{4}{c}{\textsc{tbabs*diskbb}} \\ 
\cline{2-4}\cline{6-9}\\
ObsID (Date)       & $\Gamma$           &   $\chi^2_{\nu}$/d.o.f. & PL Flux && $T_{in}$ &  Disk Norm &  $\chi^2_{\nu}$/d.o.f. &  Disk Flux  \\
                   &                    &   \textbf{or} (W-stat)            &($10^{-14}$ \ergcms)&& (keV)&($10^{-3}$)&\textbf{or} (W-stat)&($10^{-14}$ \ergcms)\\ 
\hline
352 (2000-07-29)$\dagger$& $\dagger$ & $\leq$2.1 &$\dagger$&&$\dagger$&$\dagger$&$\dagger$&$\dagger$\\\
2707 (2002-07-06)& 2.79 ($^{+1.6}_{-1.0}$)&1.4 $^{+0.9}_{-0.6}$&(148.79/172)&& 0.7 $^{+0.6}_{-0.4}$& $\leq$1.8 & (149.62/172) &0.9$^{+0.4}_{-0.4}$\\
3717 (2002-07-05) $\dagger$& $\dagger$ & $\leq$2.0 &$\dagger$&&$\dagger$&$\dagger$&$\dagger$&$\dagger$\\\
5826 (2005-03-03) & 1.7 ($\pm0.3$)& 1.09/22 & 5.5 ($\pm 1.1$) && 1.1 $^{+0.4}_{-0.3}$ & 1.5 ($^{+2.3}_{-1.0}$)&1.03/33& 2.3 ($\pm 1.0$) \\
5827 (2005-05-05)& 2.5 ($\pm 0.3$) &1.29/33 &2.2 ($\pm 0.3$) && 0.5 $^{+0.2}_{-0.1}$ &20.1 ($^{+46.0}_{-14.5}$) & 1.52/33 &1.6 ($\pm 0.3$)\\
5828 (2005-11-17)& 2.0 ($^{+0.6}_{-0.5}$)& (82.55/90) & 6.7($\pm 0.8$)&& 0.6 ($^{+0.6}_{-0.2}$)& $\leq$ 5.3& (83.00/90)&1.6 ($^{+0.8}_{-0.6}$)\\
6186 (2005-01-31) & 2.1 ($\pm 0.2$)& 1.61/16& 1.7  ($\pm0.9$)&&0.8 ($\pm0.1$) &7.7 ($^{+7.0}_{-3.6}$)& 1.48/16&5.1 ($\pm0.7$)\\
7210 (2005-11-16)& 2.5 ($^{+1.1}_{-0.9}$)& 0.26/4& 3.0  ($^{+0.4}_{-0.2}$) && 0.5 ($^{+0.4}_{-0.2}$) & $\leq$16.7 & 0.25/4 &2.0 ($\pm0.6$) \\
7211 (2005-11-16)& 2.1 ($\pm 0.4$)& (56.59/56)& 5.0($^{+1.3}_{-1.1}$) && 0.8 ($^{+0.3}_{-0.2}$)&5.1 ($^{+8.6}_{3.4}$)& (52.40/56) &4.0 ($^{+1.0}_{-0.8}$)\\
11783 (2010-04-13)  $\dagger$& $\dagger$  &$\dagger$& $\leq$1.0&&$\dagger$&$\dagger$&$\dagger$&$\dagger$\\
\hline
\end{tabular}
\end{table*}

\begin{table*}
\centering
    \caption{\textit{Chandra} Fit Parameters and Fluxes (0.5-8 keV) for spectral best fit single-component models,  \texttt{tbabs*pegpwrlw} and\texttt{tbabs*diskbb} for J04-GCULX. Hydrogen column density ($N_H$) frozen to 2.54 $\times 10^{20}$cm$^{-2}$. Lower count observations fit with W-stat have their statistics presented in parentheses. All fluxes shown are unabsorbed. Values marked with $\dagger$ are upper-limits based on the count rate of a non-detection (see Table \ref{uplims}), values marked with - were detections but could not be fit with the \texttt{diskbb} model.}
\label{ulxfit-J04GCULX}
\begin{tabular}{ccccccccc}
\hline
\hline
\multicolumn{9}{c}{J04-GCULX}\\ 
\hline
                    & \multicolumn{3}{c}{\textsc{tbabs*pegpwrlw}} && \multicolumn{4}{c}{\textsc{tbabs*diskbb}} \\ 
\cline{2-4}\cline{6-9}\\
ObsID (Date)       & $\Gamma$           &   $\chi^2_{\nu}$/d.o.f. & PL Flux && $T_{in}$ &  Disk Norm &  $\chi^2_{\nu}$/d.o.f. &  Disk Flux  \\
                   &                    &   \textbf{or} (W-stat)            &($10^{-14}$ \ergcms)&& (keV)&($10^{-3}$)&\textbf{or} (W-stat)&($10^{-14}$ \ergcms)\\ 
\hline
352 (2000-07-29)  & 2.4 ($\pm 0.4$)& 1.23/28&3.4 ($\pm 0.9$)&& 0.4 ($\pm 0.1$) & 109.0 ($^{+163.4}_{-69.4}$)& 1.26/28&2.5 ($\pm0.05$)\\
2707 (2002-07-06) & 2.1 ($\pm 0.2$) &1.09/59& 2.7 ($\pm 0.5$) && 0.6 ($\pm 0.1$) & 10.1 ($\pm 1.1$) & 0.89/59&2.2 ($\pm 0.4$)\\
3717 (2002-07-05) & 1.5 ($\pm$0.7)& 0.42/14&6.2 ($^{+3.8}_{-2.3}$)&&1.1 ($^{+2.8}_{-0.5}$)&$\leq$1.5 &0.51/14&4.3 ($^{+6.2}_{-1.6}$)\\
5826 (2005-03-03) & 1.0 ($\pm 0.3$)&2.46/45&3.6 ($\pm 0.7$)&&2.1 ($^{+1.8}_{-0.6}$)&$\leq$0.2& 2.34/45&3.2($\pm 0.7$)  \\
5827 (2005-05-05) & 2.3 ($\pm 0.2$)&0.80/60&3.6($\pm0.5$)&&0.8$\pm0.2$)&4.8 ($^{+5.8}_{-2.6}$)&0.86/60&2.8 ($\pm 0.4$)\\
5828 (2005-11-17) & 2.3 ($^{+1.9}_{-1.6}$)& 0.64/10&2.3$^{+1.6}_{-1.0}$&&1.2 ($^{+1.2}_{-0.4}$)& $\leq$ 4.7& (126.98/122)&2.9 ($^{+1.0}_{-0.9}$)\\
6186 (2005-01-31) &$\dagger$& $\dagger$ & $\leq$ 3.2&$\dagger$&$\dagger$&$\dagger$&$\dagger$\\
7210 (2005-11-16) & 3.0 ($^{+1.0}_{-0.8}$)& 0.183/8&3.7 ($^{+1.4}_{-1.2}$)&&0.5& $\leq$38.9&0.37/8&2.5 ($\pm0.8$) \\
7211 (2005-11-16) & 1.5 ($\pm 0.9$)&(81.81/70)&3.8($^{+2.3}_{-1.6}$)&&-&-&-&-\\
18781 (2016-02-24)& 2.7 ($\pm$0.8)&1.12/19&3.6 ($\pm1.0$)&&0.7 ($^{+0.6}_{-0.4}$)&$\leq$7.0&1.42/19&2.4 ($\pm0.8$)\\
18782 (2016-02-26)& 2.8 ($\pm 0.8$)&1.49/12&2.9 ($\pm 1.0$)&& 0.5 ($^{0.3}_{-0.2}$)& $\leq$ 22.4&2.0 ($\pm0.7$)\\
18783 (2016-04-20)& 1.6 ($^{+1.1}_{-1.6}$)&(106.96/111) &2.2 ($^{+1.4}_{-0.9}$)&&-&-&-&-\\
18836 (2016-04-28)& 2.4 ($^{+0.8}_{-0.9}$)&1.34/12&3.4 ($\pm1.0$)&&0.5 ($^{1.2}_{-0.3}$)&$\leq$15.3&1.63/12&2.2($\pm 0.7$) \\
18838 (2016-05-28)& 2.9 ($\pm0.7$)&0.64/21&2.3 ($\pm0.7$)&&0.4 ($^{+0.2}_{0.1}$)&$\leq$ 96.7&0.61/21&1.8 ($\pm0.6$)  \\
18856 (2016-06-12)& 1.9 ($^{+1.0}_{-2.1}$) & (104.28/106)&2.1 ($\pm1.2$)&&-&-&-&-  \\
\hline
\end{tabular}
\end{table*}

\begin{table*}
\centering
    \caption{\textit{Chandra} Fit Parameters and Fluxes (0.5-8 keV) for spectral best fit single-component models,  \texttt{tbabs*pegpwrlw} and\texttt{tbabs*diskbb} for M87-GCULX1. Hydrogen column density ($N_H$) frozen to 2.54 $\times 10^{20}$cm$^{-2}$. Lower count observations fit with W-stat have their statistics presented in parentheses. All fluxes shown are unabsorbed. Values marked with - were detections but could not be fit with the \texttt{diskbb} model.}
\begin{tabular}{ccccccccc}
\hline
\hline
\multicolumn{9}{c}{M87-GCULX1}\\ 
\hline
                    & \multicolumn{3}{c}{\textsc{tbabs*pegpwrlw}} && \multicolumn{4}{c}{\textsc{tbabs*diskbb}} \\ 
\cline{2-4}\cline{6-9}\\
ObsID (Date)       & $\Gamma$           &   $\chi^2_{\nu}$/d.o.f. & PL Flux && $T_{in}$ &  Disk Norm &  $\chi^2_{\nu}$/d.o.f. &  Disk Flux  \\
                   &                    &   \textbf{or} (W-stat)            &($10^{-14}$ \ergcms)&& (keV)&($10^{-3}$)&\textbf{or} (W-stat)&($10^{-14}$ \ergcms)\\ 
\hline
241 (2000-07-17) &3.0 ($^{+0.9}_{-0.7}$) & (21.40/27)& 4.2 ($^{+1.6}_{-1.3}$)&&0.4($^{+0.2}_{-0.1}$)&101.0 ($^{+656.1}_{-110.4}$)& (23.44/27)&3.4 ($^{+1.3}_{-1.1}$)  \\ 
352 (2000-07-29) & 1.8 ($\pm 0.1$)&1.74/22&7.5($\pm 1.1$)&& 0.6 ($\pm 0.1$)&20.9 ($^{+12.1}_{-7.8}$)&0.90/22&5.0 ($\pm$0.6)\\
2707 (2002-07-06)& 2.1 ($\pm 0.1$) &2.42/43&4.9 ($\pm0.4$) &&0.6 ($\pm 0.1$)& 19.7 ($^{+10.3}_{-6.6}$)&3.10/43&3.8($\pm0.3$) \\
3717 (2002-07-05) &1.9 ($\pm$ 0.3)&0.69/8 & 5.7 ($^{+1.3}_{-1.1}$)&&0.7 ($^{+0.2}_{-0.1}$&11.3($\pm 89.0$)& 0.76/8& 4.0 ($\pm$ 0.8) \\
5826 (2005-03-03) & 2.2 ($\pm 0.3$)& 1.51/11&4.5 ($\pm 0.6$)&& 0.7 ($\pm 0.2$)&7.4($^{+11.6}_{-4.4}$)&2.06/11&3.3 ($\pm 0.5$) \\
5827 (2005-05-05)&1.9 ($\pm 0.1$)&1.40/32&4.5($\pm0.4$)&&0.9 ($\pm 0.1$)&3.2 ($\pm 1.8$)&1.57/32 & 3.9($\pm 0.4$) \\
5828 (2005-11-17)& 2.7 ($^{+0.8}_{-0.6}$)&0.85/8& 5.0 ($\pm 0.9$)&&0.4 ($^{+0.3}_{-0.1}$)&$\leq$54.3& 3.32/8&2.8 ($\pm 0.6$)\\
6186 (2005-01-31 )& 2.1 ($\pm 0.2$)&1.33/12&4.0 ($\pm 0.8$)&& 0.8 ($\pm 0.2$)&6.1 ($^{+6.9}_{-3.1}$)&1.36/12 &4.5($\pm 0.6$) \\
7210 (2005-11-16 )& 2.3 ($\pm 0.7$)& 2.32/6 & 5.8 ($^{+1.5}_{-1.1}$) &&0.6 ($^{+0.3}_{-0.2}$)& $\leq$13.8 & 3.50/6 & 3.6 ($^{+0.9}_{-7.2}$)  \\
7211 (2005-11-16 )& 2.3 ($\pm 0.4$)&(30.99/63)&5.3($\pm 1.1$)&&0.8 ($\pm 0.3$)& 6.0 ($\pm 4.3$)& (39.82/63)&3.6 ($\pm 1.0$)\\
18781 (2016-02-24)& 2.6 ($\pm 0.3$)& 2.09/12 & 4.4 ($\pm 0.8$)&& 0.5 ($\pm 0.1$)& 35.4 ($^{+46.7}_{-20.0}$)&2.08/12 & 4.2 ($\pm 0.6$)\\
18782 (2016-02-26)& 2.4 ($\pm 0.3$)&1.13/9& 4.7 ($\pm 0.9$)&& 0.6 ($\pm 0.1$)& 19.7 ($^{+32.8}_{-21.1}$)&1.67/9 & 4.2 ($\pm 0.6$)\\
\hline
\end{tabular}
\end{table*}

\begin{table*}
\centering
    \caption{\textit{Chandra} Fit Parameters and Fluxes (0.5-8 keV) for spectral best fit single-component models,  \texttt{tbabs*pegpwrlw} and\texttt{tbabs*diskbb} for M87-GCULX2. Hydrogen column density ($N_H$) frozen to 2.54 $\times 10^{20}$cm$^{-2}$. Lower count observations fit with W-stat have their statistics presented in parentheses. All fluxes shown are unabsorbed.Values marked with $\dagger$ are upper-limits based on the count rate of a non-detection (see Table \ref{uplims}), values marked with - were detections but could not be fit with the \texttt{diskbb} model. }
\label{ulxfits-GCULX2}
\begin{tabular}{ccccccccc}
\hline
\hline
\multicolumn{9}{c}{M87-GCULX2}\\ 
\hline
                    & \multicolumn{3}{c}{\textsc{tbabs*pegpwrlw}} && \multicolumn{4}{c}{\textsc{tbabs*diskbb}} \\ 
\cline{2-4}\cline{6-9}\\
ObsID (Date)       & $\Gamma$           &   $\chi^2_{\nu}$/d.o.f. & PL Flux && $T_{in}$ &  Disk Norm &  $\chi^2_{\nu}$/d.o.f. &  Disk Flux  \\
                   &                    &   \textbf{or} (W-stat)            &($10^{-14}$ \ergcms)&& (keV)&($10^{-3}$)&\textbf{or} (W-stat)&($10^{-14}$ \ergcms)\\ 
\hline
352 (2000-07-29) &1.7 ($^{+0.8}_{-0.7}$) & 0.90/8&1.2 ($\pm0.6$)&& 1.0 ($^{+2.0}_{-0.6})$&$\leq$0.5&0.98/8&1.0 ($\pm0.6$)\\
2707 (2002-07-06)& 1.6 ($\pm 0.4$)&1.07/23&1.6 ($\pm0.5$)&&1.0 ($^{+0.7}_{-0.4}$)&$\leq$0.5&1.11/23&1.1($\pm0.4$)\\
3717 (2002-07-05) & 2.1 ($^{+1.9}_{-1.1}$)&2.29/4&1.5 ($^{+1.2}_{-0.8}$)&&-&-&-&-\\
5826 (2005-03-03) &1.6 ($\pm0.4$)&1.37/21&2.1 ($\pm0.5$)&&1.5 ($^{+1.1}_{-0.5}$)&$\leq$0.2&1.57/21&1.8$ (\pm$)0.4  \\
5827 (2005-05-05)&1.3 ($\pm0.2$)&1.12/30&3.1 ($\pm0.5$)&&1.4 ($^{+0.5}_{-0.3}$)&0.3($\pm0.2$)&0.94/30&2.5 ($\pm0.5$) \\
5828 (2005-11-17)& 0.9($^{+0.6}_{-0.8}$)&(58.75/80)&2.2($^{+1.0}_{-0.8}$)&&-&-&-&-\\
6186 (2005-01-31 )& 1.6 ($\pm0.6$)&0.24/8&2.4($\pm0.7$)&&1.6($^{+1.7}_{-0.6}$)& $\leq$0.2& 0.28/8&2.1($\pm0.7$)\\
18781 (2016-02-24)&0.9 ($^{+0.6}_{-0.8}$)& (83.23/104)&2.6($^{+1.5}_{-1.8}$)&&-&-&-&-\\
18782 (2016-02-26)& 1.2 ($\pm 0.6$)& (81.61/96)&3.3 ($^{+1.4}_{-1.0}$)&&-&-&-&-\\
18783 (2016-04-20) &$\dagger$&$\dagger$&$\leq$ 2.1&&$\dagger$&$\dagger$&$\dagger$&$\dagger$ \\
18836 (2016-04-28)& 1.9 ($\pm 0.7$)&(69.16/65)&3.1($\pm1.1$)&&1.5($^{+3.3}_{-0.7}$)&$\leq$0.3&(72.76/65)&1.5($^{+0.8}_{-0.6}$)\\
18838 (2016-05-28)& 1.1($\pm1.0$)&(55.78/77)&1.9($^{+1.2}_{-0.9}$)&&-&-&-&- \\
18856 (2016-06-12)& 1.9($\pm0.7$)&(54.44/77)&2.1($^{+1.0}_{-0.8}$)&&1.1($^{+1.8}_{-0.5}$)&$\leq$0.7&(55.78/77)&1.8$^{+1.0}_{-0.7}$) \\
\hline
\end{tabular}
\end{table*}

\begin{table*}
\centering
    \caption{\textit{Chandra} Fit Parameters and Fluxes (0.5-8 keV) for spectral best fit single-component models,  \texttt{tbabs*pegpwrlw} and\texttt{tbabs*diskbb} for M87-GCULX3. Hydrogen column density ($N_H$) frozen to 2.54 $\times 10^{20}$cm$^{-2}$. Lower count observations fit with W-stat have their statistics presented in parentheses. All fluxes shown are unabsorbed.Values marked with $\dagger$ are upper-limits based on the count rate of a non-detection (see Table \ref{uplims}), values marked with - were detections but could not be fit with the \texttt{diskbb} model. }
\label{ulxfits-GCULX3}
\begin{tabular}{ccccccccc}
\hline
\hline
\multicolumn{9}{c}{M87-GCULX3}\\ 
\hline
                    & \multicolumn{3}{c}{\textsc{tbabs*pegpwrlw}} && \multicolumn{4}{c}{\textsc{tbabs*diskbb}} \\ 
\cline{2-4}\cline{6-9}\\
ObsID (Date)       & $\Gamma$           &   $\chi^2_{\nu}$/d.o.f. & PL Flux && $T_{in}$ &  Disk Norm &  $\chi^2_{\nu}$/d.o.f. &  Disk Flux  \\
                   &                    &   \textbf{or} (W-stat)            &($10^{-14}$ \ergcms)&& (keV)&($10^{-3}$)&\textbf{or} (W-stat)&($10^{-14}$ \ergcms)\\ 
\hline
352 (2000-07-29) & $\dagger$&$\dagger$&$\leq$3.7&&$\dagger$&$\dagger$&$\dagger$&$\dagger$\\
2707 (2002-07-06)&1.1 ($\pm 0.4$)&(156.64/173)&1.9($\pm 0.5$)&&-&-&-&-\\
3717 (2002-07-05) & 1.9 ($^{+1.0}_{-0.9}$)&(49.64/69)&1.2($^{+1.1}_{-0.7}$)&&0.6($^{+1.0}_{-0.3}$)&$\leq3.9$&(48.73/69)&0.8($^{+0.7}_{-0.4}$)\\
5826 (2005-03-03) & 1.6($\pm0.2$)& (191.14/203&3.4($\pm0.5$)&&-&-&-&-\\
5827 (2005-05-05)&1.3($\pm0.2)$ &(183.55/243)&4.4($\pm0.5$)&&2.1($^{+0.7}_{-0.4}$)&0.1($\pm 0.1$)&(187.27/243)&4.1($\pm0.5$)\\
5828 (2005-11-17)& 1.7($\pm0.4$)&(77.52/97)&3.8($^{+1.0}_{-0.9}$)&&1.3($^{+0.5}_{-0.3}$)&$\leq$0.7&(76.02/97&3.3($\pm0.9$)\\
6186 (2005-01-31 )& 1.3($\pm0.4$)&(126/17/146)&4.0($\pm1.0$)&&-&-&-&-\\
7210 (2005-11-16 )& 1.8($\pm0.5$)&(78.90/92)&3.0($^{+1.0}_{-0.8}$)&&-&-&-&- \\
7211 (2005-11-16 )&1.4($\pm 0.5$)&(47.41/76)&5.4($^{+2.0}_{-1.5}$)&&1.8($^{+2.8}_{-0.6}$)&$\leq$0.3&(49.22/76)&5.0($^{+2.1}_{-1.5}$) \\
18781 (2016-02-24)&2.0($\pm0.7$)&(71.05/84)&1.9($\pm0.8$)&&0.9 ($^{+0.9}_{-0.3}$)&$\leq$1.2&(72.95/84)&1.5($^{+0.7}_{-0.5}$)\\
18782 (2016-02-26)&1.4($\pm0.7$)&(74.25/76)&1.8($^{+0.8}_{-0.6})$&&-&-&-&- \\
18836 (2016-04-28)& 1.2 ($^{+0.6}_{-0.7}$)&(80.19/72)&2.0($^{+0.9}_{0.7}$)&&-&-&-&-\\
18838 (2016-05-28)& 1.3($\pm 0.6$)&(80.04/93)&2.4($^{+0.9}_{-0.7}$)&&-&-&-&- \\
18856 (2016-06-12)&1.2 ($^{+0.6}_{-0.8}$)&(51.41/62)&2.0($^{+0.8}_{-0.6}$) &&-&-&-&-\\
\hline
\end{tabular}
\end{table*}

\begin{table*}
\centering
    \caption{\textit{Chandra} Fit Parameters and Fluxes (0.5-8 keV) for spectral best fit single-component models,  \texttt{tbabs*pegpwrlw} and\texttt{tbabs*diskbb} for M87-GCULX4. Hydrogen column density ($N_H$) frozen to 2.54 $\times 10^{20}$cm$^{-2}$. Lower count observations fit with W-stat have their statistics presented in parentheses. All fluxes shown are unabsorbed.Values marked with - were detections but could not be fit with the \texttt{diskbb} model. }
\label{ulxfits-GCULX4}
\begin{tabular}{ccccccccc}
\hline
\hline
\multicolumn{9}{c}{M87-GCULX4}\\ 
\hline
                    & \multicolumn{3}{c}{\textsc{tbabs*pegpwrlw}} && \multicolumn{4}{c}{\textsc{tbabs*diskbb}} \\ 
\cline{2-4}\cline{6-9}\\
ObsID (Date)       & $\Gamma$           &   $\chi^2_{\nu}$/d.o.f. & PL Flux && $T_{in}$ &  Disk Norm &  $\chi^2_{\nu}$/d.o.f. &  Disk Flux  \\
                   &                    &   \textbf{or} (W-stat)            &($10^{-14}$ \ergcms)&& (keV)&($10^{-3}$)&\textbf{or} (W-stat)&($10^{-14}$ \ergcms)\\ 
\hline
352 (2000-07-29) &1.1 ($\pm 0.41$)&(46.37/71)&2.5 ($^{+1.1}_{-0.8}$)&&-&-&-&-\\
2707 (2002-07-06)&1.3($\pm0.3$)&1.13/18&2.1($\pm0.6$)&&1.5($^{+1.2}_{-0.5}$)&$\leq$0.2&1.22/18&1.6($\pm0.6$)\\
5826 (2005-03-03) & 1.6($\pm0.5$)&1.07/7&1.3($\pm0.3$)&&1.7($^{+1.7}_{-0.6}$)&$\leq$0.1&1.57/7&1.2($\pm0.4$)\\
5827 (2005-05-05)&1.2($\pm0.4$)&0.69/7&1.7($\pm0.5$)&&2.1($^{+4.6}_{-0.8}$)&$\leq0.1$&0.72/7&1.5($\pm0.6$)\\
5828 (2005-11-17)&2.1($^{+0.7}_{-0.6}$)&(41.79/43)&1.7($^{+0.6}_{-0.5}$)&&1.0($^{+0.7}_{-0.4}$)&$\leq$ 0.8&(43.85/43)&1.3($^{+0.6}_{0.4}$) \\
7210 (2005-11-16 )& 2.0 ($^{+0.7}_{-0.6}$)&(33.68/31)&1.5($^{+0.7}_{-0.5}$)&&1.0($^{+0.9}_{0.4}$)&$\leq$0.7&(35.77/31)&1.3($^{+0.6}_{-0.5}$) \\
\hline
\end{tabular}
\end{table*}

%\begin{table*}
%\centering
 %   \caption{\textit{Chandra} Fit Parameters and Fluxes (0.5-8 keV) for spectral best fit single-component models,  \texttt{tbabs*pegpwrlw} and\texttt{tbabs*diskbb} for M87-GCULX5. Hydrogen column density ($N_H$) frozen to 2.54 $\times 10^{20}$cm$^{-2}$. Lower count observations fit with W-stat have their statistics presented in parentheses. All fluxes shown are unabsorbed. }
%\label{ulxfits-GCULX5}
%\begin{tabular}{|lclclclc|}
%\hline
%\hline
%\toprule
 % &    & & M87-GCULX5\ \\ \hline
 % &       \multicolumn{3}{c}{\textsc{tbabs*pegpwrlw}} & \multicolumn{4}{|c}{\textsc{tbabs*diskbb}} \\ \hline
%\midrule
% ObsID (Date)  & $\Gamma$ &   $\chi^2_{\nu}$/d.o.f. & PL Flux & $T_{in}$ &  Disk Norm &  $\chi^2_{\nu}$/d.o.f. &  Disk Flux  \\
%&& \textbf{or} (W-stat)&($10^{-14}$ $\frac{\mathrm{erg}}{ \mathrm{cm}^{2} s}$)& (keV)&($10^{-3}$)& \textbf{or} (W-stat)&($10^{-14}$ $\frac{\mathrm{erg}}{ \mathrm{cm}^{2} s}$)\\ \hline  \hline

%4007 (2003-11-21) &1.57($\pm0.27$)&1.17/13&6.87($\pm1.50$)&0.96($^{+0.35}_{-0.22}$)&2.78($^{+4.60}_{-1.90}$)&1.27/13&4.51($^{+1.19}_{-0.89}$)\\

%\\ \hline
%\end{tabular}
%\label{ulx5fits}

\begin{table}
\begin{tabular}{cccc}
\hline \hline
Source      & ObsID  & Count Rate            & Flux Upper Limits     \\ 
            &         & (ct/s)                & erg s$^{-1}$cm$^{-2}$ \\
\hline
L18-GCULX   & 352    & 4.1$\times 10^{-3}$  & 2.09$\times 10^{-14}$ \\
M87-GCULX3  & 352    & 4.7$\times 10^{-3}$   & 3.71$\times 10^{-14}$ \\
L18-GCULX   & 3717   & 3.9$\times 10^{-3}$   & 1.99$\times 10^{-14}$ \\
J04-GCULX   & 6186   & 5.11$\times 10^{-3}$   & 3.23$\times 10^{-14}$ \\
L18-GCULX   & 11783  & 1.6$ \times 10^{-3}$  &1.06$\times 10^{-14}$ \\
M87-GCULX2  & 18783  & 1.8$\times 10^{-3}$   & 2.14$\times 10^{-14}$ \\ 
\hline
\end{tabular}
\caption{Background subtracted upper limit count rates and upper flux limit estimates for GCULXs in the 0.5-8keV band.}
\label{uplims}
\end{table}

%\end{table*}

\subsection{X-Ray Variability}
We searched for intra-observational variability by  extracting background subtracted source lightcurves using \textsc{ciao}'s \texttt{dmextract} in the longest observations of the sources (over 100ks per observation). The majority of the sources revealed no clear intra-observational variability. Two of these sources, J04-GCULX and M87-GCULX1 were also studied by \citealt{Foster13}, and showed no periodic variability on the scale of weeks to years. 

\begin{figure*}
\includegraphics[scale=0.75]{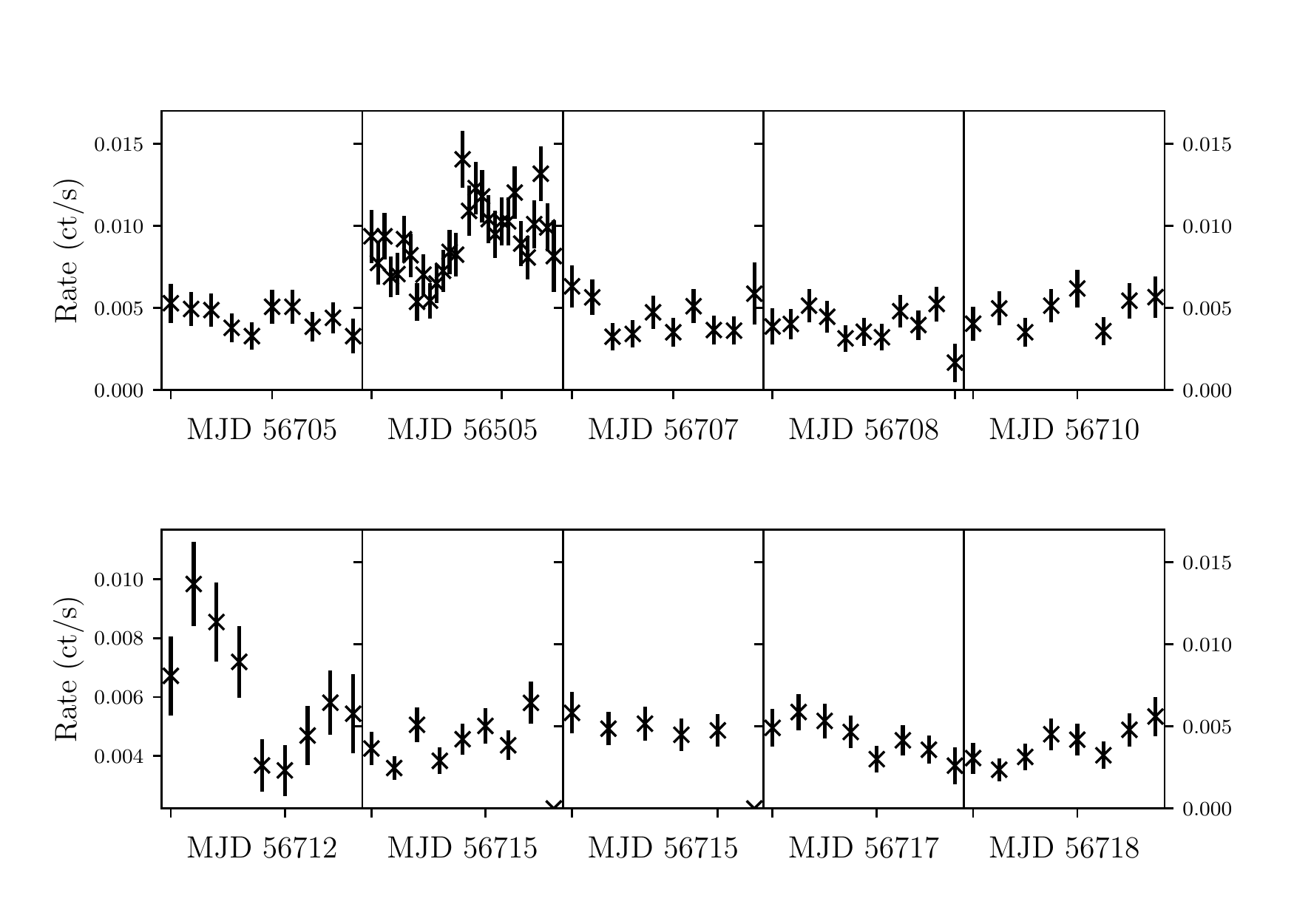}

\caption{Background-subtracted light curves from SC302 across all observations, binned by 5ks in the 0.3-10 keV band. The source shows significant variability in two of the observations, and is one of the few GCULXs to do so. } 
\label{fig:lcsc302}
\end{figure*}

The lightcurves of the majority of these sources also showed no clear short-term changes on the scale of hours (these lightcurves are plotted in Appendix \ref{timing}). The exception to this is SC302, which showed strong intra-observational variability in some (but not all) of the extracted light curves. We extracted the light curves from all observations of this source (Figure \ref{fig:lcsc302}).  

 We used FTOOLS lcstats\footnote{\url{https://heasarc.gsfc.nasa.gov/docs/xanadu/xronos/help/lcstats.html}} to quantify the variability in each of the observations, and found that the RMS fractional variation of every observation was a 3$\sigma$ upper limit, with the exception of SC302 during four observations. These values are presented in Table \ref{argelfraster}.  
\begin{table}
\caption{RMS values for individual observations of SC302.}
\label{argelfraster}
\begin{tabular}{ll}
ObsID & RMS \\ \hline
15179 & 0.32 ($\pm$   0.11)\\
15180 &0.24 ($\pm$   0.08)\\
16590 &0.37 ($\pm$ 0.13)\\
16591 &0.39 ($\pm$  0.16) \\\hline
\end{tabular}
\end{table}

\section{Results}
\label{sec:res}
Using the wealth of archival \textit{Chandra} data available for M87 and its outlying regions, we identified 7 ultraluminous X-ray sources which are associated with M87's globular clusters. Of these, one cluster (SC302) is spectroscopically confirmed (matched to Strader et al in prep, see also \citealt{caldwell14}) and the rest are photometric GC candidates (matched to sources from catalogs from \citealt{2004ApJ...613..279J, 2016MNRAS.455..820O}). 

We extracted and fit the X-ray spectra of these sources across $\sim$ 16 years worth of extant data, using a single absorbed power-law model and a single absorbed black-body disk model. We cannot differentiate between the models for the sources over most of the observations, as most of the sources had a similar $\chi^2$/d.o.f. to each other for each model. SC302 and M87-GCULX1 were better fit by an absorbed multi-colored disk and an absorbed power-law respectively for some observations, and equally well in the rest.

\subsection{ Power-Law Model Fits}
\begin{figure*}

\includegraphics[scale=0.5]{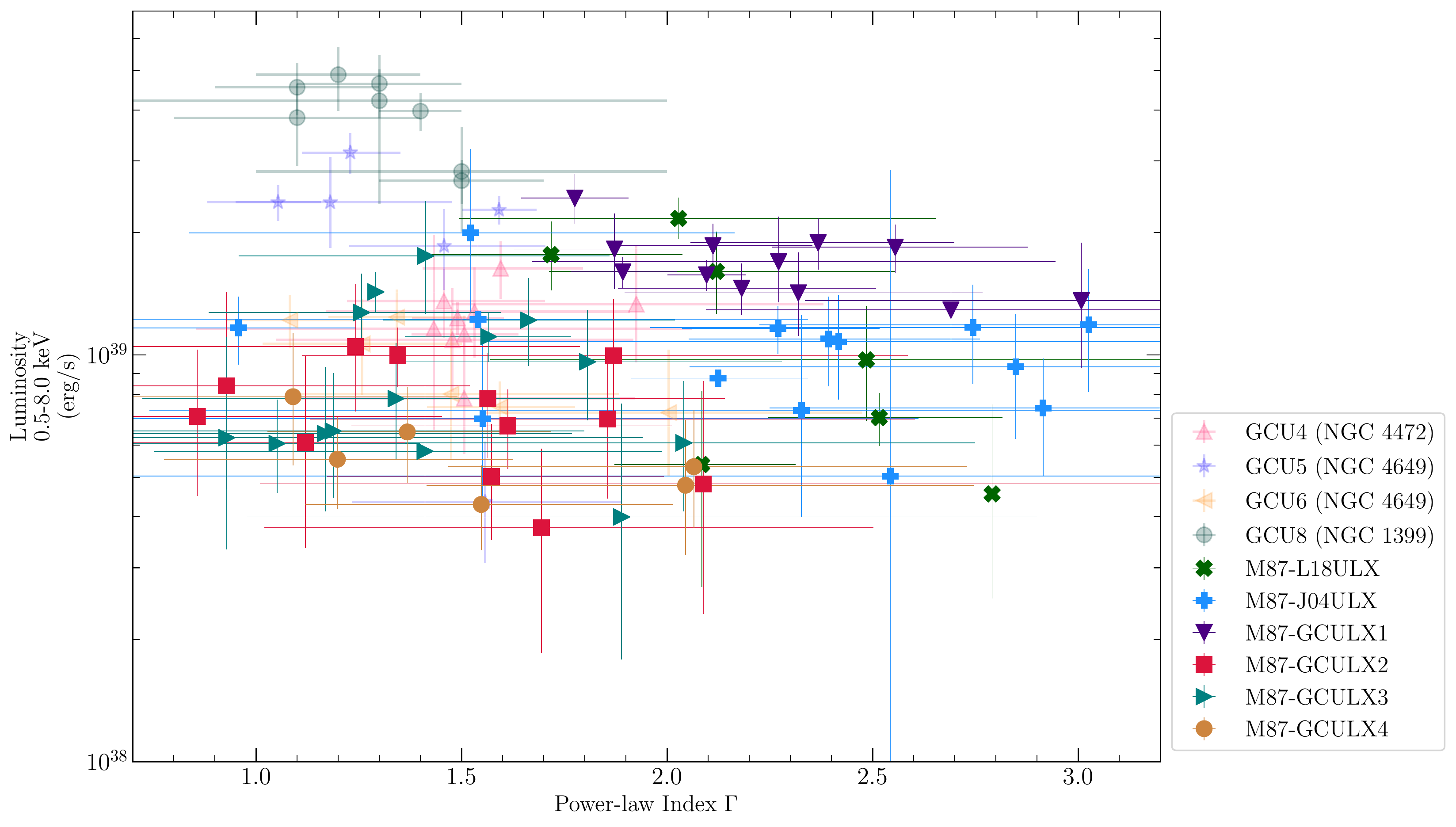}

\caption{$L_X$ versus $\Gamma$ for M87's newly discovered GC ULX population compared to fit parameters for other known GC ULXs \citep{dage19a}.  } 
\label{fig:plfits}
\end{figure*}

The best fit power-law indices of the sources in the M87 sample have been plotted against their unabsorbed X-ray luminosities in Figure \ref{fig:plfits}, along with the same values for the previously studied GC ULXs in \citealt{dage19a}. To evaluate any potential correlations between X-ray luminosity (in log scale) and power-law photon index or disk blackbody inner temperature, we used Pearson's r and Spearman's rank. We note that all data points have non-negligible uncertainties. Thus, we used bootstrap sampling (random draws for each data point based on the uncertainties) and calculated correlation coefficient for each randomized sample and evaluated the significance of correlation for each source based on the final distribution of correlation coefficients. We found that there is no statistical evidence for clear correlation between the best-fit X-ray parameters and the X-ray luminosity, except perhaps a $\sim$ 1$\sigma$ suggestion of correlation in M87-GCULX3. These sources seem to show a similar clustering to the previously studied sources, except for GCU8 in NGC~1399, which appears to be much brighter and harder than the rest of the sample. J04-GCULX shows an interestingly wide range of best fit power-law indices. 
 
\subsection{ Blackbody Disk Model Fits}
\begin{figure*}

\includegraphics[scale=0.5]{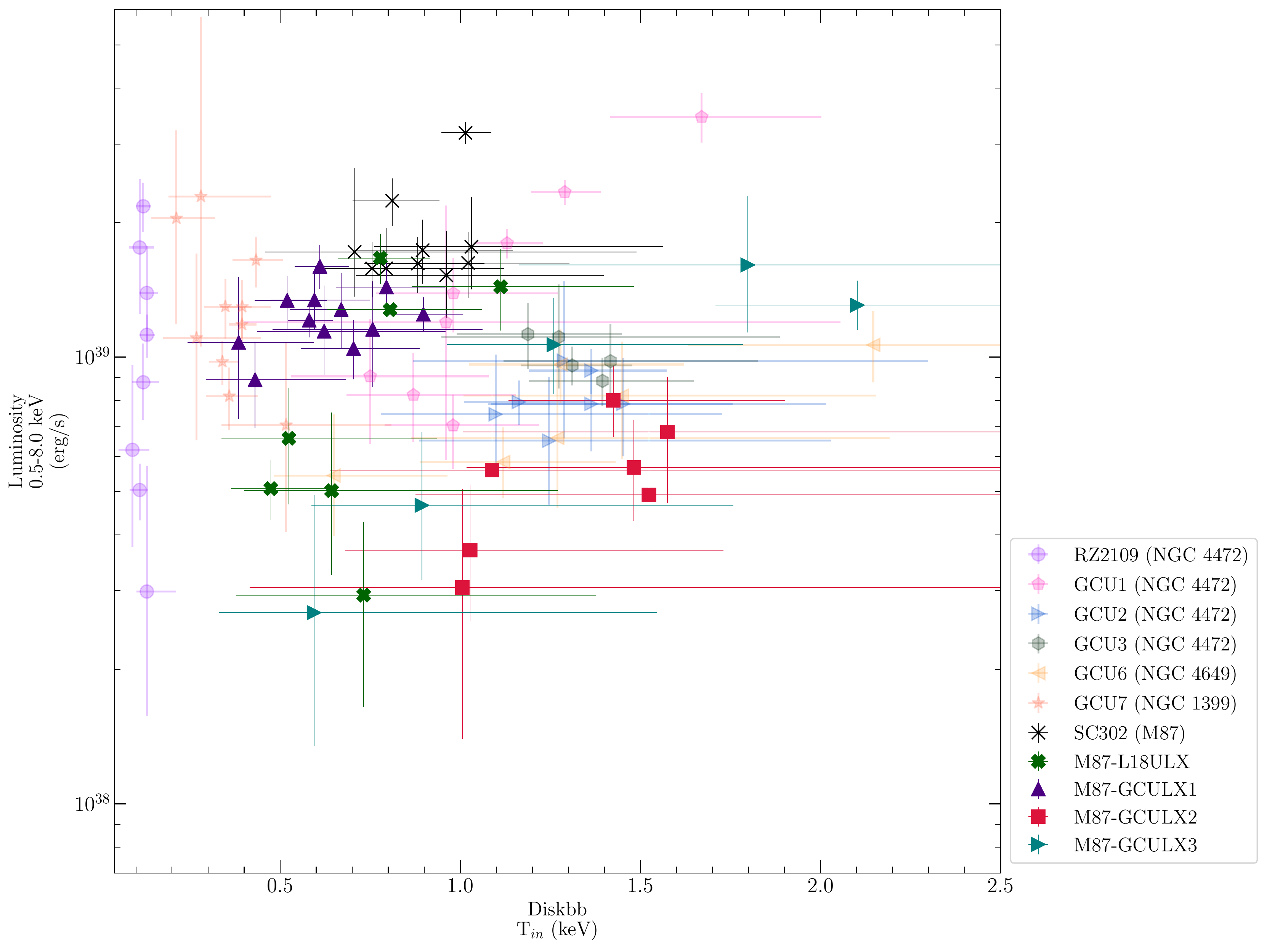}

\caption{$L_X$ versus $T_{in}$ for M87's newly discovered GC ULX population compared to fit parameters for other known GC ULXs \citep{dage19a}. } 
\label{fig:disks}
\end{figure*}

In Figure \ref{fig:disks}, we plot the best fit inner disk temperature (kT$_{in}$) for each observation versus the corresponding X-ray luminosity, with the same values for the larger GC ULX population plotted in the background for comparison. None of the M87 GC ULX sources appear to behave like GCU7 in NGC~1399 or RZ2109 in NGC~4472, which have very self-consistent values kT$_{in}$ and a wide range of luminosities, which highlights how rare these two sources are. 

%Many of the sources, particularly SC302, M87-GCULX2, and J04-GCULX, appear to follow a similar trend to GCU1 in NGC~4472. This trend shows kT increasing with $L_X$. The exact nature of this trend is difficult to constrain due to the large error bars, but blah blah blah statistical thing that Arash volunteered himself for by commenting on it. This trend can be likened to the sources presented in Kajava and Poutanen 2009, including many of the fainter galactic BH sources such as LMC X-3 and GRO J1655-40 This behaviour appears to be common to GC ULXs, but not global

%Arash: I made it to here... will continue later
\subsection{Short and Long Term-X-Ray Variability of Sources}

Lightcurves from this study, as well as the sources in \citealt{dage19a} suggest that GC ULXs do not typically vary within an observation.  RZ2109 \citep{maccarone07}, the GC ULX in NGC~1399 which faded below the detection limit of any subsequent observations after 2003 \citep{shih10}, and now SC302 seem to be the exceptions to this rule with their strong intra-observational variability.

\begin{figure*}
    \centering
    \begin{minipage}{0.45\textwidth}
        \centering
        \includegraphics[width=3.75in]{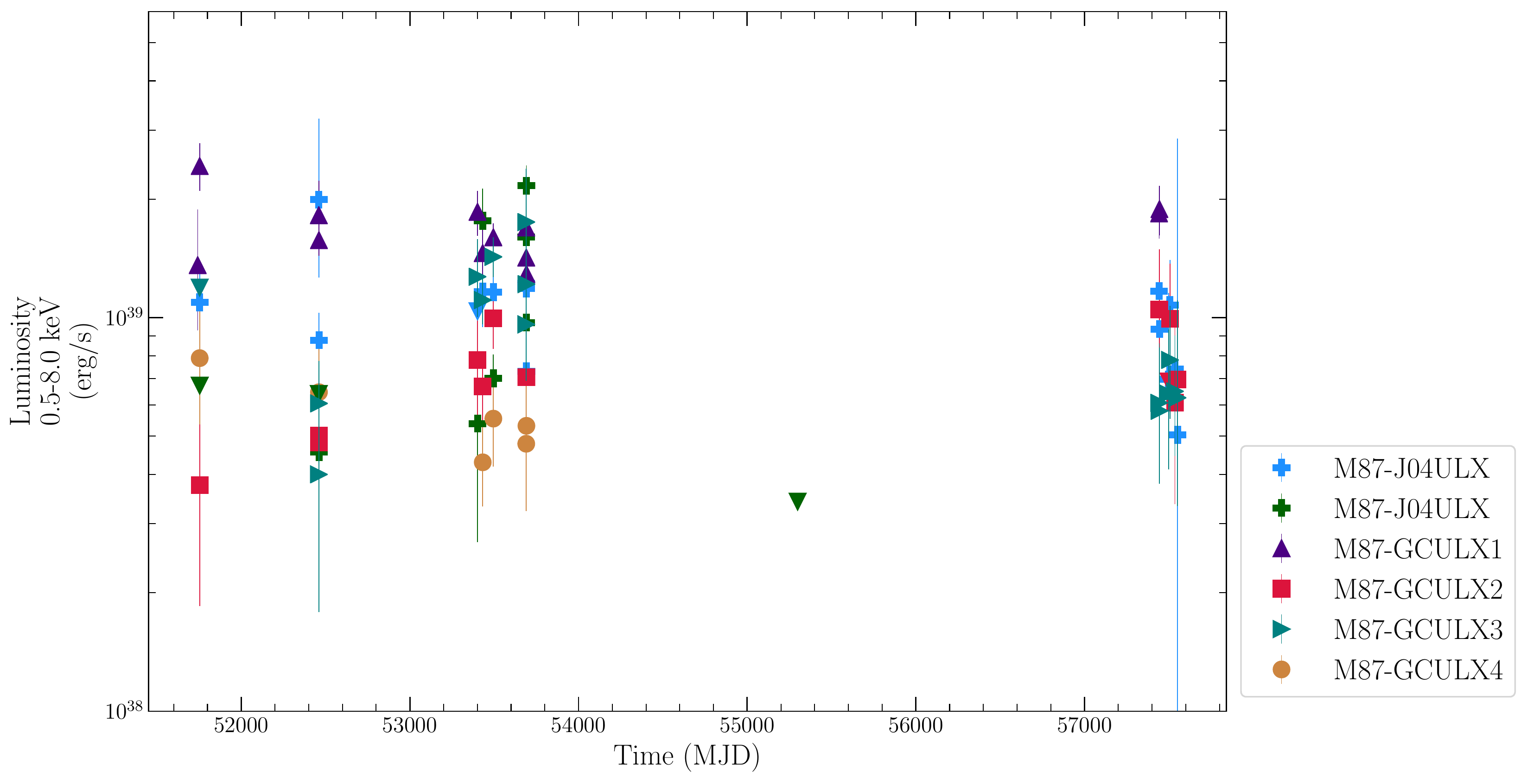} % first figure itself
    \end{minipage}\hfill
    \begin{minipage}{0.45\textwidth}
        \centering
        \includegraphics[width=3.75in]{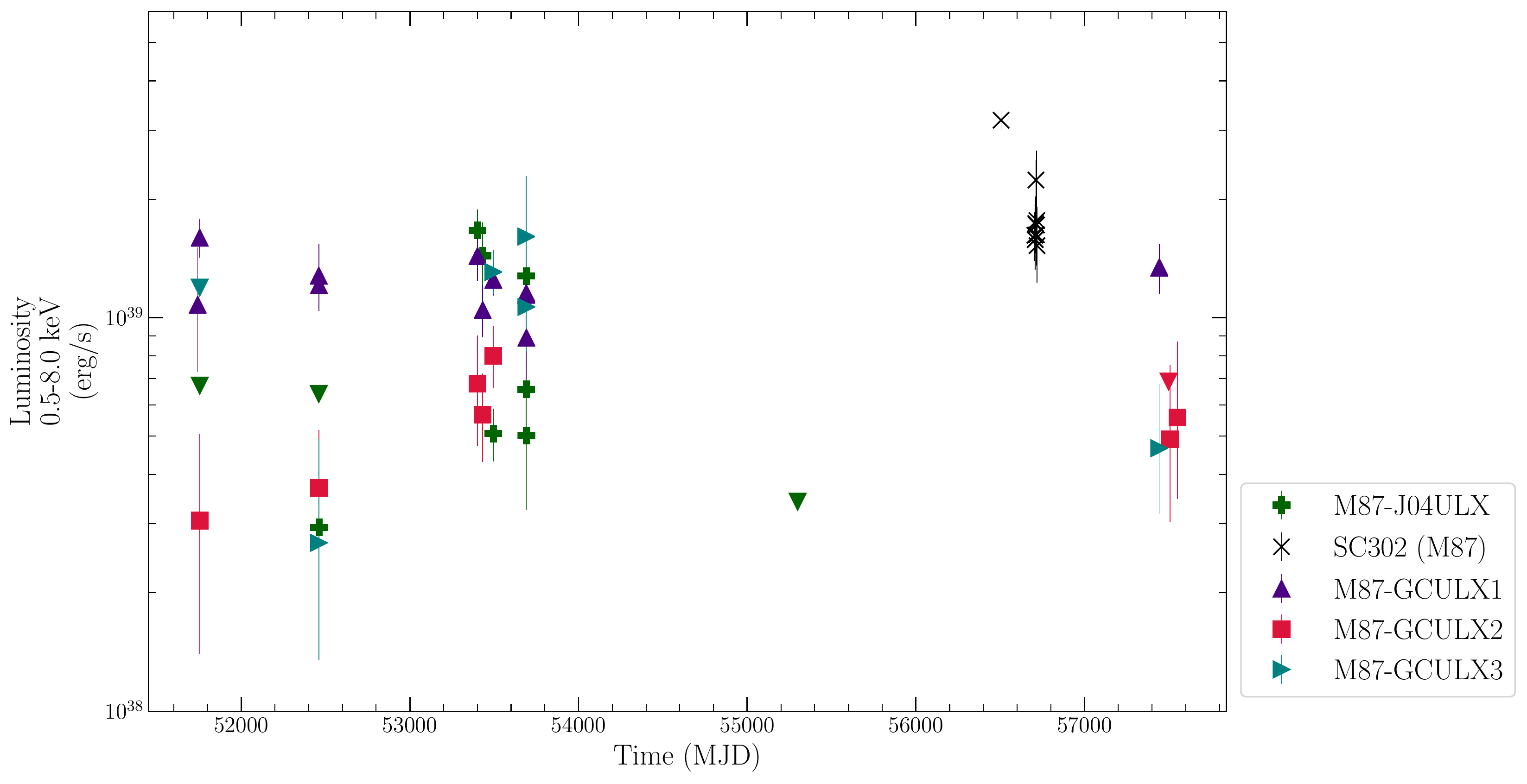} % second figure itself
    \end{minipage}
        \caption{$L_X$ versus MJD for M87's newly discovered GC ULX population, with $L_X$ calculations based off of best fit \textsc{pegpwrlw} fluxes (left) and best fit \textsc{diskbb} fluxes (right). }
        \label{fig:pltime}

\end{figure*}

Figure \ref{fig:pltime} shows the best fit unabsorbed X-ray luminosity values for each source versus the MJD of the observation, along with upper limits. %from Table \ref{uplims}.
SC302 does not appear to vary greatly in $L_X$ across observations, which is interesting because the source appears to vary within a factor of two during some observations. SC302's long-term variability was only probed over the course of around a year, and further monitoring could reveal interesting long-term variability. It is of note that none of these sources rose above 4$\times10^{39}$ erg/s. 
J04-GCULX shows some evidence of variability, as it was observed bright in all observations except for being an upper limit/non-detection in ObsID 6186 (2005-01-31). M87-GCULX3 was initially observed as an upper limit with a luminosity below $2\times 10^{38}$ erg/s in ObsID 352 (2000-07-29), and brightening in subsequent observations.  M87-GCULX2 is observed  bright until ObsID 18782 (2016-02-26) and fades  in ObsID 18783 (2016-04-20), but rebrightens for ObsIDs 18836-18856 (2016-04-28, 2016-05-28, 2016-06-12).

\subsection{Comparison of X-ray and Optical Parameters}
We use the K-S test \footnote{\url{https://docs.scipy.org/doc/scipy-0.14.0/reference/generated/scipy.stats.ks\_2samp.html}} to compare magnitudes to the greater cluster population (as identified by \citep{jordan09} ). The K-S statistic probability is 0.57 (p value=0.006), which implies that the magnitudes of the host cluster population and the overall cluster population are very different. The Anderson-Darling test statistic is 7.00 (significance level=0.001), which is in agreement with the result of the K-S test. Therefore, the presences of ULXs in the clusters are not evenly distributed across different cluster parameters, but tend to reside in the most luminous (and also the more massive) clusters. 

A K-S test of the metallicity of the ULX hosting clusters compared to the clusters identified in \citep{jordan09} gave a  probability of 0.38 (p value=0.15). The Anderson-Darling test statistic is 0.68, with a significance level equal to 0.17.  These numbers indicate that one cannot meaningfully separate the ULX hosting clusters from the rest of the sample based on their color.  Previous work by \citealt{kundu02, sarazin03} have indicated that metal rich globular clusters are three times as likely to host low mass X-ray binaries. Many subsequent analyses have firmly established this correlation for X-ray binaries in GCs. But this data set strongly suggests that the affinity for metal rich GCs does not extend to the most luminous subset of XRBs which make up the ULX population. This tentative disagreement may be due to the physics behind the formation and evolution of ULX systems in globular clusters compared to the less luminous X-ray binaries, although it is worth noting that the majority of these clusters are not spectroscopically confirmed, and that SC302 is also in the mass range of a stripped nucleus (see Section \ref{sec:summ}). 

\begin{figure*}
 \centering
    \begin{minipage}{0.45\textwidth}
        \centering
        \includegraphics[width=3.75in]{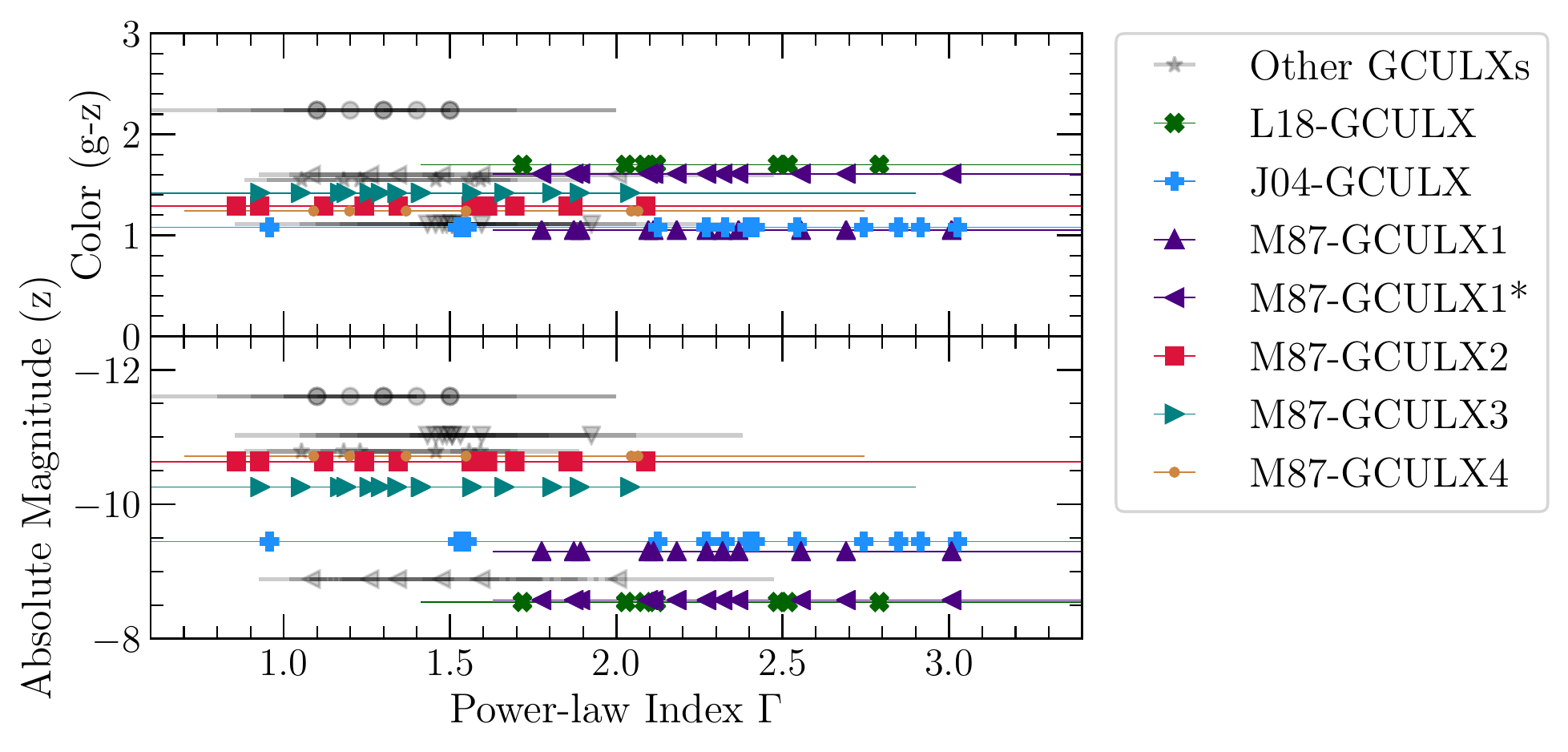} % first figure itself
    \end{minipage}\hfill
    \begin{minipage}{0.45\textwidth}
        \centering
        \includegraphics[width=3.75in]{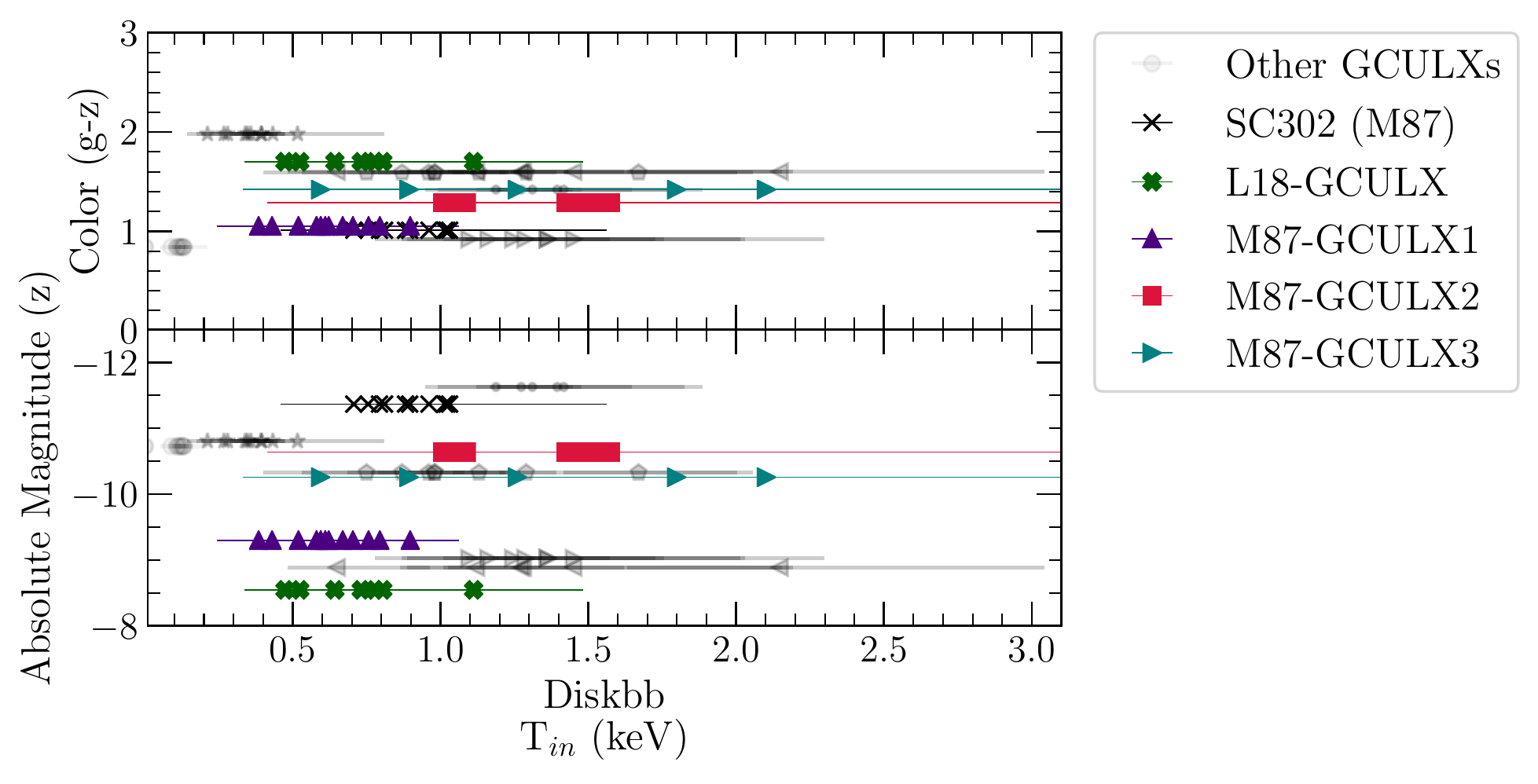} % second figure itself
    \end{minipage}
        \caption{Color and magnitude versus best fit power-law index (left) and kT (right) for M87 GC ULXs compared to previous GC ULXs \citealt{dage19a}.}
\label{fig:xray_z_gz}
\end{figure*}

Figure \ref{fig:xray_z_gz} shows the best fit X-ray parameters for both models plotted against the absolute $z$ magnitude and color ($g-z$) of the source's cluster companions. The  color distribution of the M87 cluster hosts is more concentrated, and neither very red, nor very blue, but previous ULX hosting clusters have been observed to be both extremely red and extremely blue. This suggests that the M87 sources are more typical hosts of GC ULXs, with RZ2109, GCU7 and GCU8 being the outliers in this strange population of bright X-ray sources. 

A study of M31's globular clusters by \citealt{Trudolyubov04} show that the most luminous X-ray sources ($\sim 10^{38}$ erg/s ) are associated with the most metal poor clusters. However, this does not seem to be true for M87's ultraluminous population, as the clusters values for (g-z) all ranged above 1.0--although other GCULXs such as RZ2109 have (g-z) values below 1.0.  \citealt{2007ApJ...662..525K} suggest that for most GCs with a high $L_X$, that luminosity is coming in majority from only one low mass X-ray binary (LMXB) source, and that the only clusters that are likely to host many bright LMXB sources are the very metal rich clusters, which M87's ULX population also does not fall into. 

\subsection{Summary of Sources}
\label{sec:summ}
In the following subsections we present a summary and discussion of the seven sources analysed in this paper.
\subsubsection{CXOUJ122959.82123812.33 (SC302)}
SC302 is in the Virgo Cold Front, and 19' from M87's galaxy centre. It is the only spectroscopically confirmed GC in this sample of GC hosts, with $z$=19.76 and (g-z)=1.01.  However,  the cluster mass is 3.33$\times 10^{6}$ $M_\odot$, which is in the range of either a GC or stripped nucleus. This source is the brightest optical cluster in this sample, and also the most metal poor (although it is not as metal poor as some of the other known GCULXs such as RZ2109).  
The source's best-fit power-law index based on the stacked spectra  is $\Gamma$=2.07 and the \textsc{diskbb} kT= 0.92 keV. $\chi ^2$ statistics of the deep spectrum suggest that this source is better fit by the \textsc{diskbb} model. 

While there are many observations of this source, they have been taken over approximately one year in span, so we do not have a good idea of the long-term variability of this source; however, it shows curious short-term variability on the scale of hours from its extracted lightcurves. It varies both in $L_X$ and in its spectral parameter in a similar fashion to GCU1 in NGC~4472 \citep{maccarone11, dage19a}. It is also the brightest X-ray source in this sample, with its peak X-ray luminosity reaching close to $4\times 10^{39}$ erg/s. 
\subsubsection{CXOUJ123054.97+122438.51 (L18-GCULX)}
L18-GCULX's host cluster is both the faintest cluster and most metal rich in the sample, with $z$=22.58, $(g-z)$=1.70, and cluster mass 2.53$\times 10^{5}$ $M_\odot$. $\Gamma$=2.03 ($\pm$0.11) and kT= 0.81 ($\pm$0.09) keV. This source shows a strong variation in  $L_X$ over the course of many observations, but little variability within an observation. 

\subsubsection{CXOUJ123049.24+122334.52 (J04-GCULX) }
 J04-GCULX is the closest source to the galaxy centre, and the GC has $z$=21.68 and (g-z)=1.08, with the cluster mass 5.70$\times 10^{5}$ $M_\odot$. Overall, the source has a softer power-law index. The best-fit spectral parameters of the stacked spectrum are  $\Gamma$= 2.21 ($\pm$ 0.09)  and kT= 0.74 ($\pm$ 0.07) keV. The fit statistics of the stacked spectrum do not delineate between either model. Note that the high background near the galaxy center affected background subtraction of the source's spectrum, specifically with the lower counts. 
 
 Long-term, the source appears to vary moderately near $1\times 10^{39}$ erg/s, with some brighter and fainter excursions over the full dataset. However, within an observation,  shows no clear variability  \citep{Foster13}, see Appendix \ref{timing}.

\subsubsection{CXOUJ123047.12+122416.06 (M87-GCULX1) }
 M87-GCULX was matched to two possible host clusters which have $z$=21.93 (cluster mass = 5.04$\times 10^{5}$ $M_\odot$) or 22.55 (estimated cluster mass=2.60$\times 10^{5}$ $M_\odot$) and (g-z)=1.05 or 1.61.  Its stacked X-ray spectrum appears to be better fit by  \texttt{tbabs*(diskbb+pegpwrlw)}, with kT$\sim$0.56 keV and $\Gamma$ $\sim$1.95. However, we were unable to fit the individual observations with the two component model without either of the components having large uncertainties consistent with zero. 
 
 Long-term, over the course of the observations, its average X-ray luminosity tends to vary between 1-3$\times 10^{39}$ erg/s. Within an observation, similar to  J04-GCULX, it also shows no clear variability within the observations \citep{Foster13}, see Appendix \ref{timing}.

\subsubsection{CXOUJ123050.12+122301.14 (M87-GCULX2)}
 The properties of the host cluster of M87-GCULX2 are $z$=20.49, (g-z)=1.29, and an estimated cluster mass of 1.73 $\times 10^{6}$ $M_\odot$. There was no clear delineation between models for this source. The best-fit parameters of the stacked spectrum are  $\Gamma$ = 1.52($\pm$ 0.11) and kT = 1.44 ($\pm$ 0.21).The fit statistics of the stacked spectrum do not delineate between either model. Note that the high background near the galaxy center affected background subtraction of the source's spectrum, specifically with the lower counts.
 
 This source shows interesting variability. It is one of the fainter ultraluminous X-ray sources, only reaching about $10^{39}$ erg/s in a few observations. It also shows variability on timescales of $\sim$ days; it was observed near $10^{39}$  erg/s on 2016-02-26, but was a non-detection on 2016-04-20, but 8 days later in the next observation, it reached $10^{39}$  erg/s again, and remained bright in the following observations.  It is quite likely that the source shows high amplitude variability, which could be confirmed by a monitoring campaign on this region.

\subsubsection{CXOUJ123049.97+122400.11 (M87-GCULX3)}
 M87-GCULX3's cluster has optical properties of $z$=20.87 and (g-z)=1.42. The estimated cluster mass is 1.22 $\times 10^{6}$ $M_\odot$.The  X-ray spectrum of this source needed to be binned by 1 and fit with W-statistics in all observations, and is one of the faintest X-ray sources in this sample, only reaching about $10^{39}$ erg/s in one observation when fit with \texttt{tbabs*pegpwrlw}, so the source is not even securely classified as ultraluminous. Its best fit spectral parameters based on the stacked spectrum are $\Gamma$=1.36 ($\pm$ 0.09)  and kT= 1.90 ($\pm$ 0.28)keV. The source shows no clear short term variability, but these fits are not well constrained in the individual observations due to the faintness of the source. The fit statistics of the stacked spectrum do not delineate between either model.
 
 Unlike the rest of the sources in this sample, M87-GCULX3 was below $10^{39}$ erg/s in almost all of the observations. It likely has a  different physical mechanism than the rest of the sources in this sample, and may possibly be comparable to Z track neutron star sources which show a roughly similar luminosity variability \citep[e.g.][]{2018AstL...44..593N}. Some candidate Z sources have been identified in globular clusters such as Terzan-5 and GCs in M31 \citep{2003A&A...411..553B, 2012ApJ...748...82L}, as well as extragalatic globular clusters associated with NGC 3115 (although none of these were observed to reach $10^{39}$ erg/s) \citep{Lin15}.

In a study of globular cluster X-ray sources in NGC~3379, \citealt{Brassington10} found a source with X-ray luminosities on the order of $10^{38}$ erg/s which reached close to $10^{39}$ erg/s in an observation, and was best fit with $\Gamma$ = 1.85, which may also be similar to M87-GCULX3.

\subsubsection{CXOUJ123041.77+122440.16 (M87-GCULX4)}
 M87-GCULX4's host cluster is $z$=20.41, (g-z)=1.24 and an estimated cluster mass of 1.86 $\times 10^{6}$ $M_\odot$. This source is one of the hardest in the sample, with $\Gamma$= 1.37 ($\pm 0.15$)   kT= 1.86 ($^{+0.56}_{-0.38}$), based on the stacked spectrum, with the fit statistics favouring the power-law model as the best fit model. This source's spectrum was also affected due to the the high background near the galaxy center. 
 
 The variability in this source is interesting in the long-term, as it was observed turning on in the first two observations, and rising to just above $10^{39}$ erg/s for many of the early observations. The cluster of observations shows the source clustered just below $10^{39}$ erg/s. The lightcurves do not appear to show any short term variability. 
%\subsubsection{CXOUJ123017.77+122545.15 (M87-GCULX5)}
% The cluster hosting M87-ULX5 is $z$=21.78 and (g-z)=1.58. This cluster is the most metal rich in the sample presented in this paper, although it is by no means the most metal rich cluster to host a ULX. There is only one observation of this source because it was in a separate field than the other sources. M87-GCULX-5 is one of the brighter measurements in $L_X$ with a power-law index near 1.6, and  kT
%$\sim$ 1.0 keV. The light-curve for this source shows no inter-observational variability, but we recommend further monitoring of this source to determine if it varies long-term.  

\section{Conclusions}
\label{sec:conc}
GC ULXs comprise a very rare population, as the chance of an individual cluster hosting a ULX is very low, indeed. Previously, only 10 had been identified, 3 associated with NGC~1399 (one of which ``turned off" in 2003 and has not been observed above the observation detection limit since) \citep{shih10, irwin10, dage19a}, two associated with NGC~4649 \citep{roberts12, dage19a}, and five associated with NGC~4472 \citep{maccarone07, maccarone11, dage19a}. 

A new search of M87's bountiful GC system revealed seven new GC ULX sources, almost doubling the known sample, which is not surprising, as the size of M87's globular cluster system means that the size of the cluster sample that has been searched for ULXs has roughly doubled as well. One of these sources, SC302, is a spectroscopically confirmed GC with a measured radial velocity (Strader et al in prep, see also \citealt{caldwell14}), while the rest are photometrically selected GCs from surveys by \citep{jordan09, 2016MNRAS.455..820O}.%, which carefully model and remove the contaminant population. 
We recommend detailed optical follow-up on these sources in the future, as the sources studied in \citealt{dage19a} are postulated to show a correlation between the presence of optical emission lines and the X-ray spectral behaviour of the source.  A high-resolution spectrum would not only confirm the globular cluster nature of the host sources, but place upper limits on any optical emission coming from the source.

Many of these newly discovered sources in M87 show striking variability over the 16 years that they have been monitored, in some cases showing order of magnitude variability. However, many of these sources are also remarkably steady over the timescales probed as well, which is similar to the sample studied in \citealt{dage19a}. M87-GCULX2 is especially interesting, as it seems to vary strongly between sub-Eddington and super-Eddington timescales at least on the scale of months. %, evoking a similarity to the strong variability in RZ2109's X-ray source. 

None of the sources in the M87 sample rose above 4$\times10^{39}$ erg/s, and the overall $L_X$ of the previously studied sample suggests that these sources also do not rise above  4$\times10^{39}$ erg/s over a total observation, with the exception of GCU8 in NGC~1399. 

Interestingly, the optical properties of the cluster (i.e. colour and magnitude) again do not seem to show any correlation with the X-ray temporal or spectral properties. In fact the results of this study only serve to highlight how different the very soft sources such as RZ2109 (NGC~4472) and GCU7 (NGC~1399), or the very hard source GCU8  (NGC~1399) are different from the rest of the GC ULX population. 

We postulate that the accretors of these systems are most likely black holes. While some ULXs in young, star-forming galaxies have been identified to have neutron star primaries \citep[e.g.][]{song20}, the nature of these sources cannot inform on the nature of the accretors in the globular cluster systems due to a number of key environmental and phenomenological differences. Namely, the X-ray binaries in the star-forming regions of galaxies have massive, hydrogen rich donor stars \citep{gladstone13}, whereas only lower mass donor stars remain in the very much older globular cluster environment.  The globular cluster binaries also face a vastly more dynamic and complicated evolutionary history, with the more massive cluster component likely having exchanged multiple companions throughout its lifetime due to preferential mass exchange \citep{spitzer87}. Finally, the observational characteristics of the two populations differ greatly. The ULXs with neutron star primaries have all been observed at X-ray luminosities near $10^{40}$ erg/s \citep{2014Natur.514..202B}, whereas the GC ULXs are observed between 1-4 $\times$ $10^{39}$ erg/s. The spectra of the ULXs in star-forming regions typically consist of a soft component added to a hard component, or a broadened disk \citep{2017ARA&A..55..303K}, yet with one exception the sources in this study were best fit by single-component models. 
\section*{Data Availability Statement}
The X-ray data in this article is publicly available through the \textit{Chandra} archive\footnote{\url{https://cda.harvard.edu/chaser/}}.
\section*{Acknowledgements}
We thank the referee, Jimmy Irwin, for a helpful report which improved the quality and presentation of the results in this paper.
KCD, SEZ, and MBP acknowledge support from Chandra grant GO4-15089A. 
SEZ and MBP also acknowledge support from the NASA ADAP grant NNX15AI71G. JS acknowledges support from the Packard Foundation. KCD acknowledges Elias Aydi, Rufus Byrd, Chelsea Harris  and Jimmy Irwin for their help. 
This research has made use of data obtained from the Chandra Data Archive and the Chandra Source Catalog. We  also acknowledge use of NASA's Astrophysics Data System and Arxiv.

The following software and packages were used for analysis: \textsc{ciao}, software provided by the Chandra X-ray Center (CXC),   \textsc{heasoft} obtained from the High Energy Astrophysics Science Archive Research Center (HEASARC), a service of the Astrophysics Science Division at NASA/GSFC and of the Smithsonian Astrophysical Observatory's High Energy Astrophysics Division, SAOImage DS9, developed by Smithsonian Astrophysical Observatory,  \textsc{numpy} \citep{2011arXiv1102.1523V}, and  \textsc{matplotlib} \citep{2007CSE.....9...90H}.

\bibliographystyle{mnras}
\bibliography{m87gculx} % if your bibtex file is called example.bib

\begin{thebibliography}{}
\makeatletter
\relax
\def\mn@urlcharsother{\let\do\@makeother \do\$\do\&\do\#\do\^\do\_\do\%\do\~}
\def\mn@doi{\begingroup\mn@urlcharsother \@ifnextchar [ {\mn@doi@}
  {\mn@doi@[]}}
\def\mn@doi@[#1]#2{\def\@tempa{#1}\ifx\@tempa\@empty \href
  {http://dx.doi.org/#2} {doi:#2}\else \href {http://dx.doi.org/#2} {#1}\fi
  \endgroup}
\def\mn@eprint#1#2{\mn@eprint@#1:#2::\@nil}
\def\mn@eprint@arXiv#1{\href {http://arxiv.org/abs/#1} {{\tt arXiv:#1}}}
\def\mn@eprint@dblp#1{\href {http://dblp.uni-trier.de/rec/bibtex/#1.xml}
  {dblp:#1}}
\def\mn@eprint@#1:#2:#3:#4\@nil{\def\@tempa {#1}\def\@tempb {#2}\def\@tempc
  {#3}\ifx \@tempc \@empty \let \@tempc \@tempb \let \@tempb \@tempa \fi \ifx
  \@tempb \@empty \def\@tempb {arXiv}\fi \@ifundefined
  {mn@eprint@\@tempb}{\@tempb:\@tempc}{\expandafter \expandafter \csname
  mn@eprint@\@tempb\endcsname \expandafter{\@tempc}}}

\bibitem[\protect\citeauthoryear{{Abbott} et~al.,}{{Abbott}
  et~al.}{2016}]{abbott16}
{Abbott} B.~P.,  et~al., 2016, \mn@doi [\apjl] {10.3847/2041-8205/833/1/L1},
  \href {https://ui.adsabs.harvard.edu/abs/2016ApJ...833L...1A} {833, L1}

\bibitem[\protect\citeauthoryear{{Bachetti} et~al.,}{{Bachetti}
  et~al.}{2014}]{2014Natur.514..202B}
{Bachetti} M.,  et~al., 2014, \mn@doi [\nat] {10.1038/nature13791}, \href
  {https://ui.adsabs.harvard.edu/abs/2014Natur.514..202B} {514, 202}

\bibitem[\protect\citeauthoryear{{Barnard}, {Kolb}  \& {Osborne}}{{Barnard}
  et~al.}{2003}]{2003A&A...411..553B}
{Barnard} R.,  {Kolb} U.,   {Osborne} J.~P.,  2003, \mn@doi [\aap]
  {10.1051/0004-6361:20031513}, \href
  {https://ui.adsabs.harvard.edu/abs/2003A&A...411..553B} {411, 553}

\bibitem[\protect\citeauthoryear{{Brassington} et~al.,}{{Brassington}
  et~al.}{2010}]{Brassington10}
{Brassington} N.~J.,  et~al., 2010, \mn@doi [\apj]
  {10.1088/0004-637X/725/2/1805}, \href
  {https://ui.adsabs.harvard.edu/abs/2010ApJ...725.1805B} {725, 1805}

\bibitem[\protect\citeauthoryear{{Brightman} et~al.,}{{Brightman}
  et~al.}{2018}]{brightman18}
{Brightman} M.,  et~al., 2018, \mn@doi [Nature Astronomy]
  {10.1038/s41550-018-0391-6}, \href
  {https://ui.adsabs.harvard.edu/abs/2018NatAs...2..312B} {2, 312}

\bibitem[\protect\citeauthoryear{{Brightman} et~al.,}{{Brightman}
  et~al.}{2019}]{2019arXiv191204431B}
{Brightman} M.,  et~al., 2019, arXiv e-prints, \href
  {https://ui.adsabs.harvard.edu/abs/2019arXiv191204431B} {p. arXiv:1912.04431}

\bibitem[\protect\citeauthoryear{{Caldwell}, {Strader}, {Romanowsky}, {Brodie},
  {Moore}, {Diemand}  \& {Martizzi}}{{Caldwell} et~al.}{2014}]{caldwell14}
{Caldwell} N.,  {Strader} J.,  {Romanowsky} A.~J.,  {Brodie} J.~P.,  {Moore}
  B.,  {Diemand} J.,   {Martizzi} D.,  2014, \mn@doi [\apjl]
  {10.1088/2041-8205/787/1/L11}, \href
  {https://ui.adsabs.harvard.edu/abs/2014ApJ...787L..11C} {787, L11}

\bibitem[\protect\citeauthoryear{{Cash}}{{Cash}}{1979}]{1979ApJ...228..939C}
{Cash} W.,  1979, \mn@doi [\apj] {10.1086/156922}, \href
  {http://adsabs.harvard.edu/abs/1979ApJ...228..939C} {228, 939}

\bibitem[\protect\citeauthoryear{{Chomiuk}, {Strader}, {Maccarone},
  {Miller-Jones}, {Heinke}, {Noyola}, {Seth}  \& {Ransom}}{{Chomiuk}
  et~al.}{2013}]{2013ApJ...777...69C}
{Chomiuk} L.,  {Strader} J.,  {Maccarone} T.~J.,  {Miller-Jones} J. C.~A.,
  {Heinke} C.,  {Noyola} E.,  {Seth} A.~C.,   {Ransom} S.,  2013, \mn@doi
  [\apj] {10.1088/0004-637X/777/1/69}, \href
  {https://ui.adsabs.harvard.edu/abs/2013ApJ...777...69C} {777, 69}

\bibitem[\protect\citeauthoryear{{Dage}, {Zepf}, {Peacock}, {Bahramian},
  {Noroozi}, {Kundu}  \& {Maccarone}}{{Dage} et~al.}{2019a}]{dage19a}
{Dage} K.~C.,  {Zepf} S.~E.,  {Peacock} M.~B.,  {Bahramian} A.,  {Noroozi} O.,
  {Kundu} A.,   {Maccarone} T.~J.,  2019a, \mn@doi [\mnras]
  {10.1093/mnras/stz479}, \href
  {https://ui.adsabs.harvard.edu/abs/2019MNRAS.485.1694D} {485, 1694}

\bibitem[\protect\citeauthoryear{{Dage} et~al.,}{{Dage}
  et~al.}{2019b}]{dage19b}
{Dage} K.~C.,  et~al., 2019b, \mn@doi [\mnras] {10.1093/mnras/stz2514}, \href
  {https://ui.adsabs.harvard.edu/abs/2019MNRAS.489.4783D} {489, 4783}

\bibitem[\protect\citeauthoryear{{Dickey} \& {Lockman}}{{Dickey} \&
  {Lockman}}{1990}]{1990ARA&A..28..215D}
{Dickey} J.~M.,  {Lockman} F.~J.,  1990, \mn@doi [\araa]
  {10.1146/annurev.aa.28.090190.001243}, \href
  {https://ui.adsabs.harvard.edu/abs/1990ARA&A..28..215D} {28, 215}

\bibitem[\protect\citeauthoryear{{Fabbiano} \& {White}}{{Fabbiano} \&
  {White}}{2006}]{fabbiano06}
{Fabbiano} G.,  {White} N.~E.,  2006, {Compact stellar X-ray sources in normal
  galaxies}.
Cambridge University Press, pp 475--506

\bibitem[\protect\citeauthoryear{{Fabricant}, {Topka}, {Harnden}  \&
  {Gorenstein}}{{Fabricant} et~al.}{1978}]{1978ApJ...226L.107F}
{Fabricant} D.,  {Topka} K.,  {Harnden} F.~R. J.,   {Gorenstein} P.,  1978,
  \mn@doi [\apjl] {10.1086/182842}, \href
  {https://ui.adsabs.harvard.edu/abs/1978ApJ...226L.107F} {226, L107}

\bibitem[\protect\citeauthoryear{{Foster}, {Charles}, {Swartz}, {Misra}  \&
  {Stassun}}{{Foster} et~al.}{2013}]{Foster13}
{Foster} D.~L.,  {Charles} P.~A.,  {Swartz} D.~A.,  {Misra} R.,   {Stassun}
  K.~G.,  2013, \mn@doi [\mnras] {10.1093/mnras/stt557}, \href
  {https://ui.adsabs.harvard.edu/abs/2013MNRAS.432.1375F} {432, 1375}

\bibitem[\protect\citeauthoryear{{Fruscione} et~al.,}{{Fruscione}
  et~al.}{2006}]{2006SPIE.6270E..1VF}
{Fruscione} A.,  et~al., 2006, in Society of Photo-Optical Instrumentation
  Engineers (SPIE) Conference Series. p. 62701V, \mn@doi{10.1117/12.671760}

\bibitem[\protect\citeauthoryear{{Giersz}, {Askar}, {Wang}, {Hypki}, {Leveque}
  \& {Spurzem}}{{Giersz} et~al.}{2019}]{giersz19}
{Giersz} M.,  {Askar} A.,  {Wang} L.,  {Hypki} A.,  {Leveque} A.,   {Spurzem}
  R.,  2019, \mn@doi [\mnras] {10.1093/mnras/stz1460}, \href
  {https://ui.adsabs.harvard.edu/abs/2019MNRAS.487.2412G} {487, 2412}

\bibitem[\protect\citeauthoryear{{Giesers} et~al.,}{{Giesers}
  et~al.}{2018}]{giesers18}
{Giesers} B.,  et~al., 2018, \mn@doi [\mnras] {10.1093/mnrasl/slx203}, \href
  {https://ui.adsabs.harvard.edu/abs/2018MNRAS.475L..15G} {475, L15}

\bibitem[\protect\citeauthoryear{{Giesers} et~al.,}{{Giesers}
  et~al.}{2019}]{giesers19}
{Giesers} B.,  et~al., 2019, \mn@doi [\aap] {10.1051/0004-6361/201936203},
  \href {https://ui.adsabs.harvard.edu/abs/2019A&A...632A...3G} {632, A3}

\bibitem[\protect\citeauthoryear{{Gladstone}, {Copperwheat}, {Heinke},
  {Roberts}, {Cartwright}, {Levan}  \& {Goad}}{{Gladstone}
  et~al.}{2013}]{gladstone13}
{Gladstone} J.~C.,  {Copperwheat} C.,  {Heinke} C.~O.,  {Roberts} T.~P.,
  {Cartwright} T.~F.,  {Levan} A.~J.,   {Goad} M.~R.,  2013, \mn@doi [\apjs]
  {10.1088/0067-0049/206/2/14}, \href
  {https://ui.adsabs.harvard.edu/abs/2013ApJS..206...14G} {206, 14}

\bibitem[\protect\citeauthoryear{{Harris}}{{Harris}}{2009}]{Harris09}
{Harris} W.~E.,  2009, \mn@doi [\apj] {10.1088/0004-637X/699/1/254}, \href
  {https://ui.adsabs.harvard.edu/abs/2009ApJ...699..254H} {699, 254}

\bibitem[\protect\citeauthoryear{{Hunter}}{{Hunter}}{2007}]{2007CSE.....9...90H}
{Hunter} J.~D.,  2007, \mn@doi [Computing in Science and Engineering]
  {10.1109/MCSE.2007.55}, \href
  {http://adsabs.harvard.edu/abs/2007CSE.....9...90H} {9, 90}

\bibitem[\protect\citeauthoryear{{Irwin}, {Athey}  \& {Bregman}}{{Irwin}
  et~al.}{2003}]{irwin03}
{Irwin} J.~A.,  {Athey} A.~E.,   {Bregman} J.~N.,  2003, \mn@doi [\apj]
  {10.1086/368179}, \href
  {https://ui.adsabs.harvard.edu/abs/2003ApJ...587..356I} {587, 356}

\bibitem[\protect\citeauthoryear{{Irwin}, {Brink}, {Bregman}  \&
  {Roberts}}{{Irwin} et~al.}{2010}]{irwin10}
{Irwin} J.~A.,  {Brink} T.~G.,  {Bregman} J.~N.,   {Roberts} T.~P.,  2010,
  \mn@doi [\apjl] {10.1088/2041-8205/712/1/L1}, \href
  {https://ui.adsabs.harvard.edu/abs/2010ApJ...712L...1I} {712, L1}

\bibitem[\protect\citeauthoryear{{Irwin} et~al.,}{{Irwin}
  et~al.}{2016}]{Irwin16}
{Irwin} J.~A.,  et~al., 2016, \mn@doi [\nat] {10.1038/nature19822}, \href
  {https://ui.adsabs.harvard.edu/abs/2016Natur.538..356I} {538, 356}

\bibitem[\protect\citeauthoryear{{Jithesh}, {Anjana}  \& {Misra}}{{Jithesh}
  et~al.}{2020}]{jithesh20}
{Jithesh} V.,  {Anjana} C.,   {Misra} R.,  2020, arXiv e-prints, \href
  {https://ui.adsabs.harvard.edu/abs/2020arXiv200401796J} {p. arXiv:2004.01796}

\bibitem[\protect\citeauthoryear{{Jord{\'a}n} et~al.,}{{Jord{\'a}n}
  et~al.}{2004}]{2004ApJ...613..279J}
{Jord{\'a}n} A.,  et~al., 2004, \mn@doi [\apj] {10.1086/422545}, \href
  {https://ui.adsabs.harvard.edu/abs/2004ApJ...613..279J} {613, 279}

\bibitem[\protect\citeauthoryear{{Jord{\'a}n} et~al.,}{{Jord{\'a}n}
  et~al.}{2009}]{jordan09}
{Jord{\'a}n} A.,  et~al., 2009, \mn@doi [\apjs] {10.1088/0067-0049/180/1/54},
  \href {https://ui.adsabs.harvard.edu/abs/2009ApJS..180...54J} {180, 54}

\bibitem[\protect\citeauthoryear{{Kaaret}, {Feng}  \& {Roberts}}{{Kaaret}
  et~al.}{2017}]{2017ARA&A..55..303K}
{Kaaret} P.,  {Feng} H.,   {Roberts} T.~P.,  2017, \mn@doi [\araa]
  {10.1146/annurev-astro-091916-055259}, \href
  {https://ui.adsabs.harvard.edu/abs/2017ARA&A..55..303K} {55, 303}

\bibitem[\protect\citeauthoryear{{Kajava} \& {Poutanen}}{{Kajava} \&
  {Poutanen}}{2009}]{2009MNRAS.398.1450K}
{Kajava} J. J.~E.,  {Poutanen} J.,  2009, \mn@doi [\mnras]
  {10.1111/j.1365-2966.2009.15215.x}, \href
  {https://ui.adsabs.harvard.edu/abs/2009MNRAS.398.1450K} {398, 1450}

\bibitem[\protect\citeauthoryear{{King}}{{King}}{2008}]{king08}
{King} A.~R.,  2008, \mn@doi [\mnras] {10.1111/j.1745-3933.2008.00444.x}, \href
  {https://ui.adsabs.harvard.edu/abs/2008MNRAS.385L.113K} {385, L113}

\bibitem[\protect\citeauthoryear{{Kremer}, {Chatterjee}, {Ye}, {Rodriguez}  \&
  {Rasio}}{{Kremer} et~al.}{2019}]{kremer19}
{Kremer} K.,  {Chatterjee} S.,  {Ye} C.~S.,  {Rodriguez} C.~L.,   {Rasio}
  F.~A.,  2019, \mn@doi [\apj] {10.3847/1538-4357/aaf646}, \href
  {https://ui.adsabs.harvard.edu/abs/2019ApJ...871...38K} {871, 38}

\bibitem[\protect\citeauthoryear{{Kundu}, {Maccarone}  \& {Zepf}}{{Kundu}
  et~al.}{2002}]{kundu02}
{Kundu} A.,  {Maccarone} T.~J.,   {Zepf} S.~E.,  2002, \mn@doi [\apjl]
  {10.1086/342353}, \href
  {https://ui.adsabs.harvard.edu/abs/2002ApJ...574L...5K} {574, L5}

\bibitem[\protect\citeauthoryear{{Kundu}, {Maccarone}  \& {Zepf}}{{Kundu}
  et~al.}{2007}]{2007ApJ...662..525K}
{Kundu} A.,  {Maccarone} T.~J.,   {Zepf} S.~E.,  2007, \mn@doi [\apj]
  {10.1086/518021}, \href
  {https://ui.adsabs.harvard.edu/abs/2007ApJ...662..525K} {662, 525}

\bibitem[\protect\citeauthoryear{{Larsen}}{{Larsen}}{1999}]{larsen99}
{Larsen} S.~S.,  1999, \mn@doi [\aaps] {10.1051/aas:1999509}, \href
  {https://ui.adsabs.harvard.edu/abs/1999A&AS..139..393L} {139, 393}

\bibitem[\protect\citeauthoryear{{Lin} et~al.,}{{Lin} et~al.}{2015}]{Lin15}
{Lin} D.,  et~al., 2015, \mn@doi [\apj] {10.1088/0004-637X/808/1/19}, \href
  {https://ui.adsabs.harvard.edu/abs/2015ApJ...808...19L} {808, 19}

\bibitem[\protect\citeauthoryear{{Linares}, {Altamirano}, {Chakrabarty},
  {Cumming}  \& {Keek}}{{Linares} et~al.}{2012}]{2012ApJ...748...82L}
{Linares} M.,  {Altamirano} D.,  {Chakrabarty} D.,  {Cumming} A.,   {Keek} L.,
  2012, \mn@doi [\apj] {10.1088/0004-637X/748/2/82}, \href
  {https://ui.adsabs.harvard.edu/abs/2012ApJ...748...82L} {748, 82}

\bibitem[\protect\citeauthoryear{{Luan} et~al.,}{{Luan} et~al.}{2018}]{Luan18}
{Luan} L.,  et~al., 2018, \mn@doi [\apj] {10.3847/1538-4357/aaca94}, \href
  {https://ui.adsabs.harvard.edu/abs/2018ApJ...862...73L} {862, 73}

\bibitem[\protect\citeauthoryear{{Maccarone}, {Kundu}, {Zepf}  \&
  {Rhode}}{{Maccarone} et~al.}{2007}]{maccarone07}
{Maccarone} T.~J.,  {Kundu} A.,  {Zepf} S.~E.,   {Rhode} K.~L.,  2007, \mn@doi
  [\nat] {10.1038/nature05434}, \href
  {https://ui.adsabs.harvard.edu/abs/2007Natur.445..183M} {445, 183}

\bibitem[\protect\citeauthoryear{{Maccarone}, {Kundu}, {Zepf}  \&
  {Rhode}}{{Maccarone} et~al.}{2011}]{maccarone11}
{Maccarone} T.~J.,  {Kundu} A.,  {Zepf} S.~E.,   {Rhode} K.~L.,  2011, \mn@doi
  [\mnras] {10.1111/j.1365-2966.2010.17547.x}, \href
  {https://ui.adsabs.harvard.edu/abs/2011MNRAS.410.1655M} {410, 1655}

\bibitem[\protect\citeauthoryear{{Macri} et~al.,}{{Macri} et~al.}{1999}]{macri}
{Macri} L.~M.,  et~al., 1999, \mn@doi [\apj] {10.1086/307541}, \href
  {http://adsabs.harvard.edu/abs/1999ApJ...521..155M} {521, 155}

\bibitem[\protect\citeauthoryear{{Miller-Jones} et~al.,}{{Miller-Jones}
  et~al.}{2015}]{MillerJones15}
{Miller-Jones} J.~C.~A.,  et~al., 2015, \mn@doi [\mnras]
  {10.1093/mnras/stv1869}, \href
  {http://adsabs.harvard.edu/abs/2015MNRAS.453.3918M} {453, 3918}

\bibitem[\protect\citeauthoryear{{Mitsuda} et~al.,}{{Mitsuda}
  et~al.}{1984}]{1984PASJ...36..741M}
{Mitsuda} K.,  et~al., 1984, \pasj, \href
  {https://ui.adsabs.harvard.edu/abs/1984PASJ...36..741M} {36, 741}

\bibitem[\protect\citeauthoryear{{Nikolaeva}, {Krivonos}  \&
  {Sazonov}}{{Nikolaeva} et~al.}{2018}]{2018AstL...44..593N}
{Nikolaeva} S.~M.,  {Krivonos} R.~A.,   {Sazonov} S.~Y.,  2018, \mn@doi
  [Astronomy Letters] {10.1134/S1063773718100055}, \href
  {https://ui.adsabs.harvard.edu/abs/2018AstL...44..593N} {44, 593}

\bibitem[\protect\citeauthoryear{{Oldham} \& {Auger}}{{Oldham} \&
  {Auger}}{2016}]{2016MNRAS.455..820O}
{Oldham} L.~J.,  {Auger} M.~W.,  2016, \mn@doi [\mnras]
  {10.1093/mnras/stv2244}, \href
  {https://ui.adsabs.harvard.edu/abs/2016MNRAS.455..820O} {455, 820}

\bibitem[\protect\citeauthoryear{{Peacock} et~al.,}{{Peacock}
  et~al.}{2012}]{peacock12neb}
{Peacock} M.~B.,  et~al., 2012, \mn@doi [\apj] {10.1088/0004-637X/759/2/126},
  \href {https://ui.adsabs.harvard.edu/abs/2012ApJ...759..126P} {759, 126}

\bibitem[\protect\citeauthoryear{{Roberts} et~al.,}{{Roberts}
  et~al.}{2012}]{roberts12}
{Roberts} T.~P.,  et~al., 2012, \mn@doi [\apj] {10.1088/0004-637X/760/2/135},
  \href {https://ui.adsabs.harvard.edu/abs/2012ApJ...760..135R} {760, 135}

\bibitem[\protect\citeauthoryear{{Rodriguez}, {Zevin}, {Amaro-Seoane},
  {Chatterjee}, {Kremer}, {Rasio}  \& {Ye}}{{Rodriguez}
  et~al.}{2019}]{2019PhRvD.100d3027R}
{Rodriguez} C.~L.,  {Zevin} M.,  {Amaro-Seoane} P.,  {Chatterjee} S.,  {Kremer}
  K.,  {Rasio} F.~A.,   {Ye} C.~S.,  2019, \mn@doi [\prd]
  {10.1103/PhysRevD.100.043027}, \href
  {https://ui.adsabs.harvard.edu/abs/2019PhRvD.100d3027R} {100, 043027}

\bibitem[\protect\citeauthoryear{{Sarazin}, {Kundu}, {Irwin}, {Sivakoff},
  {Blanton}  \& {Rand all}}{{Sarazin} et~al.}{2003}]{sarazin03}
{Sarazin} C.~L.,  {Kundu} A.,  {Irwin} J.~A.,  {Sivakoff} G.~R.,  {Blanton}
  E.~L.,   {Rand all} S.~W.,  2003, \mn@doi [\apj] {10.1086/377467}, \href
  {https://ui.adsabs.harvard.edu/abs/2003ApJ...595..743S} {595, 743}

\bibitem[\protect\citeauthoryear{{Shih}, {Kundu}, {Maccarone}, {Zepf}  \&
  {Joseph}}{{Shih} et~al.}{2010}]{shih10}
{Shih} I.~C.,  {Kundu} A.,  {Maccarone} T.~J.,  {Zepf} S.~E.,   {Joseph} T.~D.,
   2010, \mn@doi [\apj] {10.1088/0004-637X/721/1/323}, \href
  {https://ui.adsabs.harvard.edu/abs/2010ApJ...721..323S} {721, 323}

\bibitem[\protect\citeauthoryear{{Shishkovsky} et~al.,}{{Shishkovsky}
  et~al.}{2018}]{shishkovsky18}
{Shishkovsky} L.,  et~al., 2018, \mn@doi [\apj] {10.3847/1538-4357/aaadb1},
  \href {https://ui.adsabs.harvard.edu/abs/2018ApJ...855...55S} {855, 55}

\bibitem[\protect\citeauthoryear{{Song}, {Walton}, {Lansbury}, {Evans},
  {Fabian}, {Earnshaw}  \& {Roberts}}{{Song} et~al.}{2020}]{song20}
{Song} X.,  {Walton} D.~J.,  {Lansbury} G.~B.,  {Evans} P.~A.,  {Fabian} A.~C.,
   {Earnshaw} H.,   {Roberts} T.~P.,  2020, \mn@doi [\mnras]
  {10.1093/mnras/stz3036}, \href
  {https://ui.adsabs.harvard.edu/abs/2020MNRAS.491.1260S} {491, 1260}

\bibitem[\protect\citeauthoryear{{Spitzer}}{{Spitzer}}{1987}]{spitzer87}
{Spitzer} L.,  1987, {Dynamical evolution of globular clusters}

\bibitem[\protect\citeauthoryear{{Steele}, {Zepf}, {Kundu}, {Maccarone},
  {Rhode}  \& {Salzer}}{{Steele} et~al.}{2011}]{steele11}
{Steele} M.~M.,  {Zepf} S.~E.,  {Kundu} A.,  {Maccarone} T.~J.,  {Rhode} K.~L.,
    {Salzer} J.~J.,  2011, \mn@doi [\apj] {10.1088/0004-637X/739/2/95}, \href
  {https://ui.adsabs.harvard.edu/abs/2011ApJ...739...95S} {739, 95}

\bibitem[\protect\citeauthoryear{{Strader} et~al.,}{{Strader}
  et~al.}{2011}]{strader11}
{Strader} J.,  et~al., 2011, \mn@doi [\apjs] {10.1088/0067-0049/197/2/33},
  \href {https://ui.adsabs.harvard.edu/abs/2011ApJS..197...33S} {197, 33}

\bibitem[\protect\citeauthoryear{{Strader}, {Chomiuk}, {Maccarone},
  {Miller-Jones}  \& {Seth}}{{Strader} et~al.}{2012}]{strader12}
{Strader} J.,  {Chomiuk} L.,  {Maccarone} T.~J.,  {Miller-Jones} J. C.~A.,
  {Seth} A.~C.,  2012, \mn@doi [\nat] {10.1038/nature11490}, \href
  {https://ui.adsabs.harvard.edu/abs/2012Natur.490...71S} {490, 71}

\bibitem[\protect\citeauthoryear{{Taylor}}{{Taylor}}{2005}]{2005ASPC..347...29T}
{Taylor} M.~B.,  2005, {TOPCAT \&amp; STIL: Starlink Table/VOTable Processing
  Software}.
Astronomical Society of the Pacific, p.~29

\bibitem[\protect\citeauthoryear{{Trudolyubov} \& {Priedhorsky}}{{Trudolyubov}
  \& {Priedhorsky}}{2004}]{Trudolyubov04}
{Trudolyubov} S.,  {Priedhorsky} W.,  2004, \mn@doi [\apj] {10.1086/425033},
  \href {https://ui.adsabs.harvard.edu/abs/2004ApJ...616..821T} {616, 821}

\bibitem[\protect\citeauthoryear{{Walton} et~al.,}{{Walton}
  et~al.}{2018}]{2018ApJ...856..128W}
{Walton} D.~J.,  et~al., 2018, \mn@doi [\apj] {10.3847/1538-4357/aab610}, \href
  {https://ui.adsabs.harvard.edu/abs/2018ApJ...856..128W} {856, 128}

\bibitem[\protect\citeauthoryear{{Zepf} et~al.,}{{Zepf} et~al.}{2008}]{zepf08}
{Zepf} S.~E.,  et~al., 2008, \mn@doi [\apjl] {10.1086/591937}, \href
  {https://ui.adsabs.harvard.edu/abs/2008ApJ...683L.139Z} {683, L139}

\bibitem[\protect\citeauthoryear{{van der Walt}, {Colbert}  \&
  {Varoquaux}}{{van der Walt} et~al.}{2011}]{2011arXiv1102.1523V}
{van der Walt} S.,  {Colbert} S.~C.,   {Varoquaux} G.,  2011, \mn@doi
  [Computing in Science and Engineering] {10.1109/MCSE.2011.37}, \href
  {https://ui.adsabs.harvard.edu/abs/2011CSE....13b..22V} {13, 22}

\makeatother
\end{thebibliography}

\clearpage
\appendix

\section{Lightcurves}
\label{timing}
Lightcurves from the longest observations of L18-GCULX, J04-GCULX, M87-GCULX1, M87-GCULX2, M87-GCULX3, and  M87-GCULX4, binned by 5ks in the 0.3-10keV band.

\begin{figure*}
    \centering
    \begin{minipage}{0.45\textwidth}
        \centering
        \includegraphics[width=3.75in]{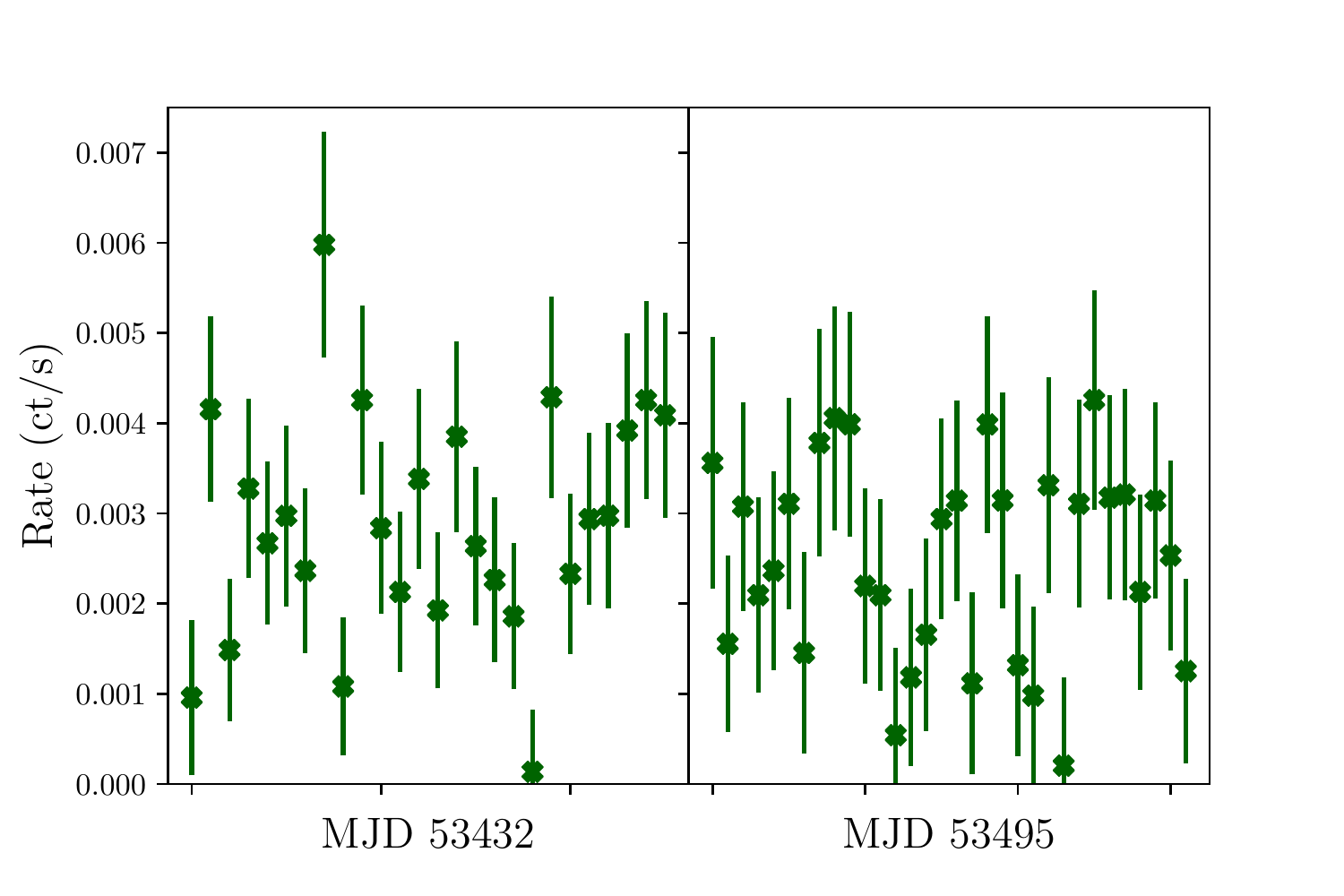} % first figure itself
    \end{minipage}\hfill
    \begin{minipage}{0.45\textwidth}
        \centering
        \includegraphics[width=3.75in]{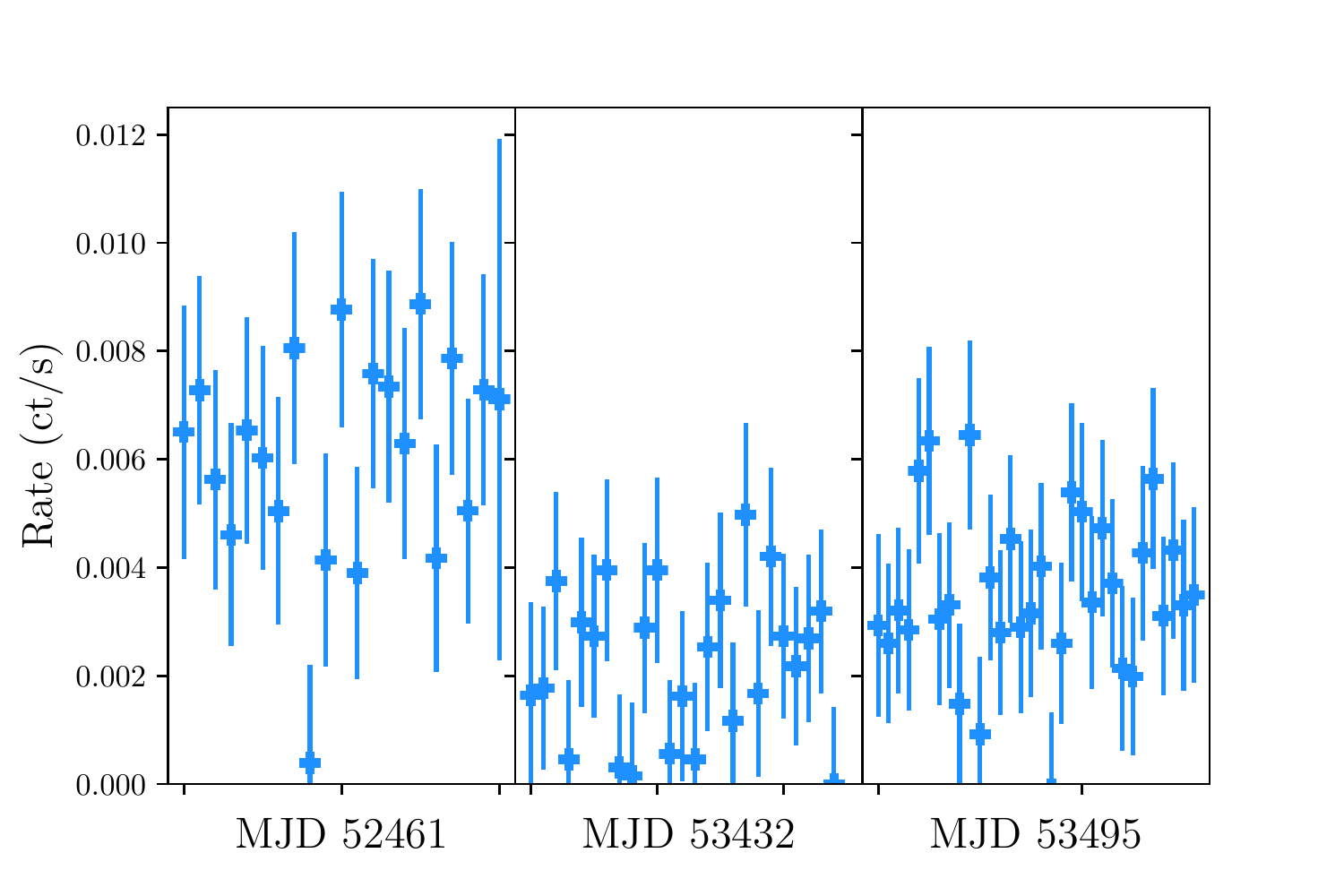} % second figure itself
    \end{minipage}
        \caption{Lightcurves from L18-GCULX (left, ObsIDs 5826 and 5827) and J04-GCULX  (right, ObsIDs 2707, 5826 and 5827 ), binned by 5ks.}
        \label{fig:firstlcs1}

\end{figure*}

\begin{figure*}
    \centering
    \begin{minipage}{0.45\textwidth}
        \centering
        \includegraphics[width=3.75in]{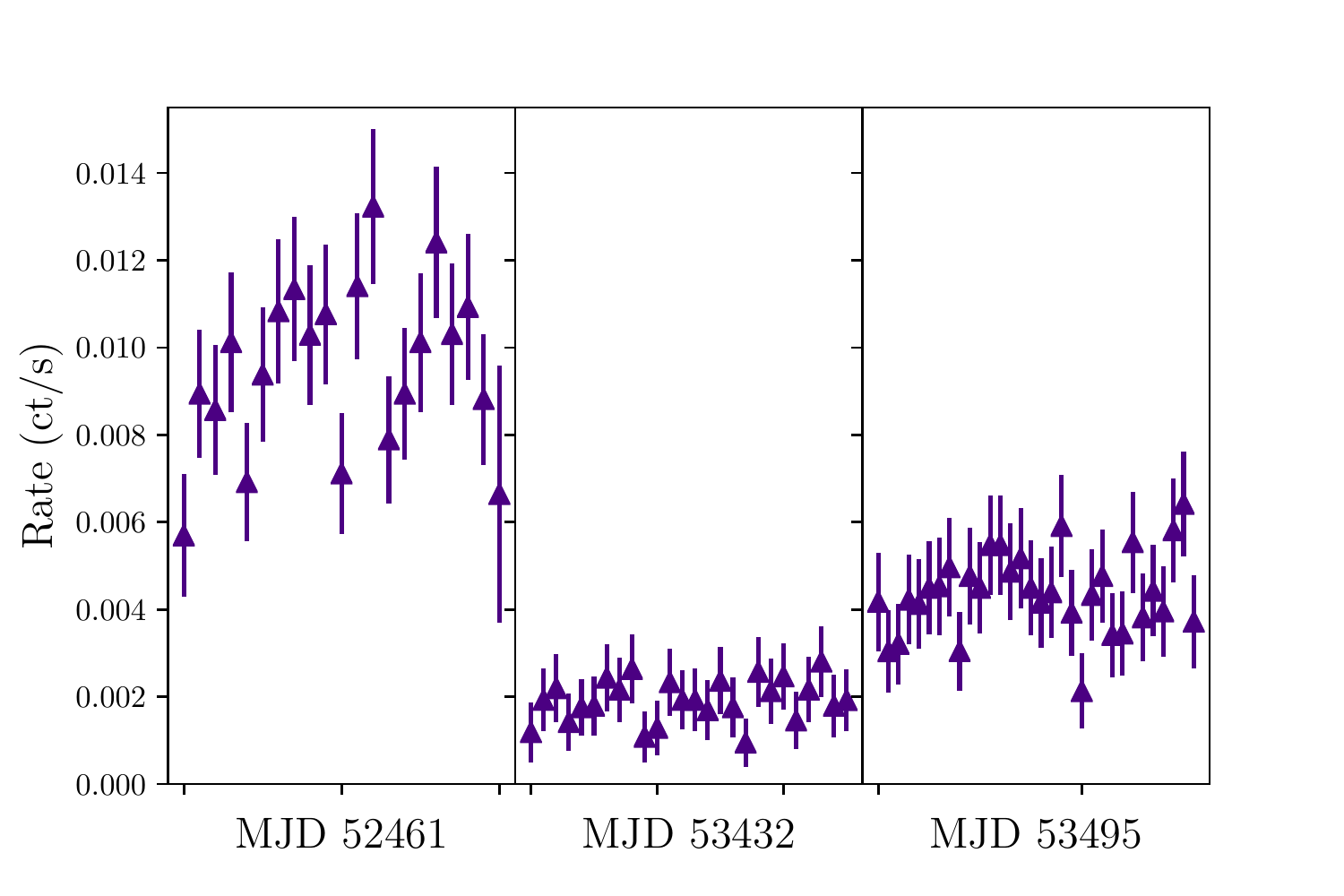} % first figure itself
    \end{minipage}\hfill
    \begin{minipage}{0.45\textwidth}
        \centering
        \includegraphics[width=3.75in]{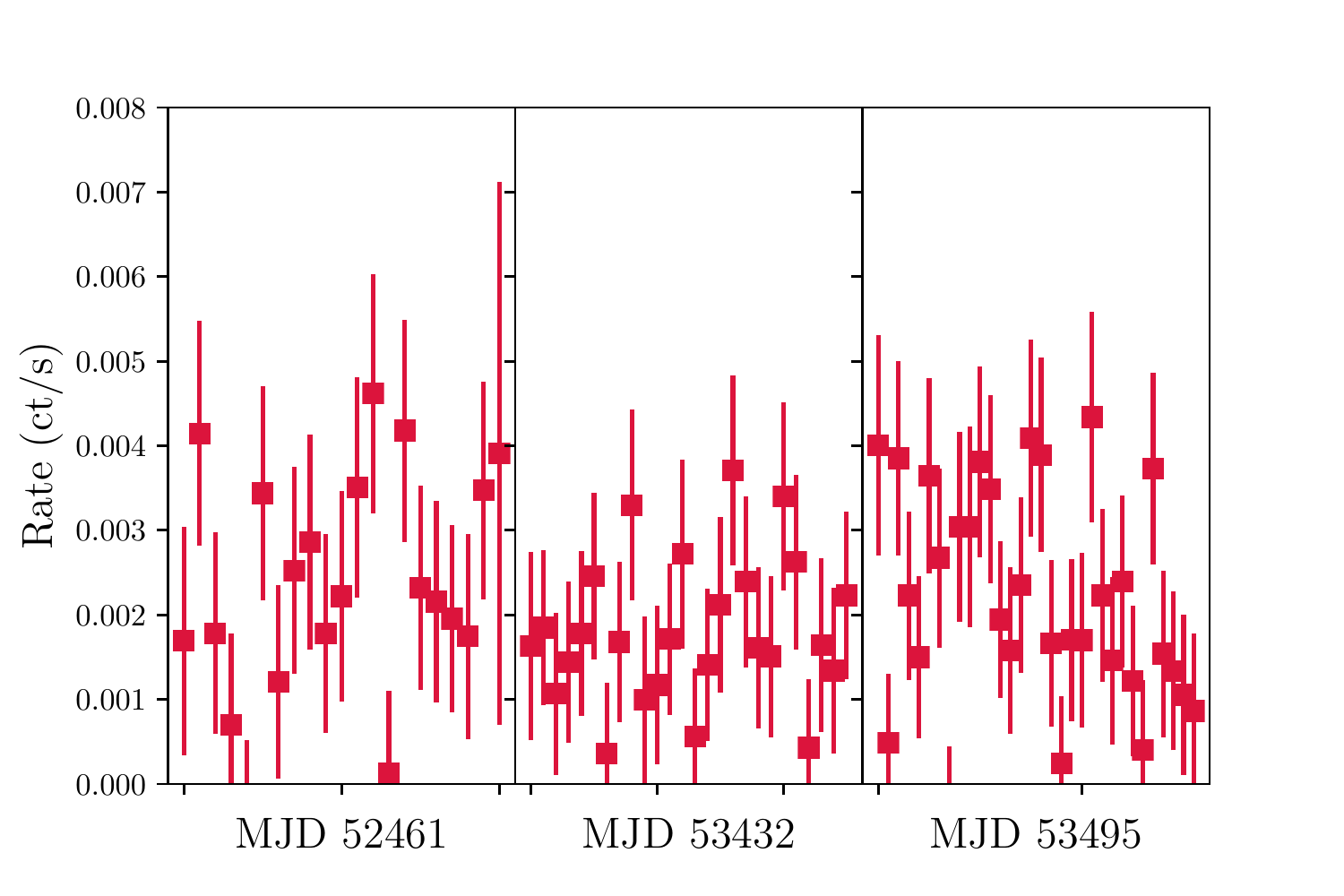} % second figure itself
    \end{minipage}
        \caption{Lightcurves from M87-GCULX1 (left) and M87-GCULX2 (right) in ObsIDs 2707, 5826 and 5827, binned by 5ks.}
        \label{fig:firstlcs2}

\end{figure*}

\begin{figure*}
    \centering
    \begin{minipage}{0.45\textwidth}
        \centering
        \includegraphics[width=3.75in]{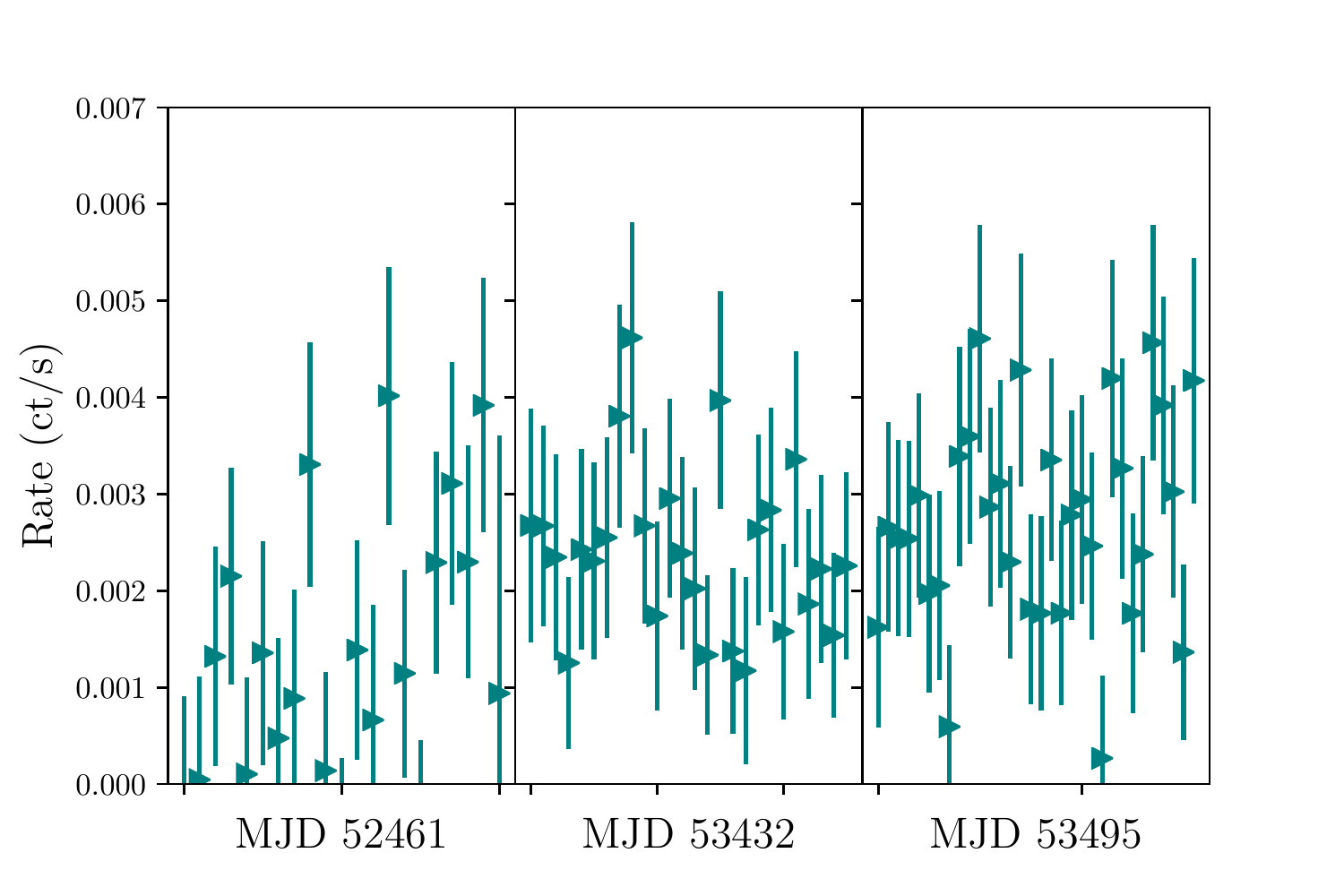} % first figure itself
    \end{minipage}\hfill
    \begin{minipage}{0.45\textwidth}
        \centering
        \includegraphics[width=3.75in]{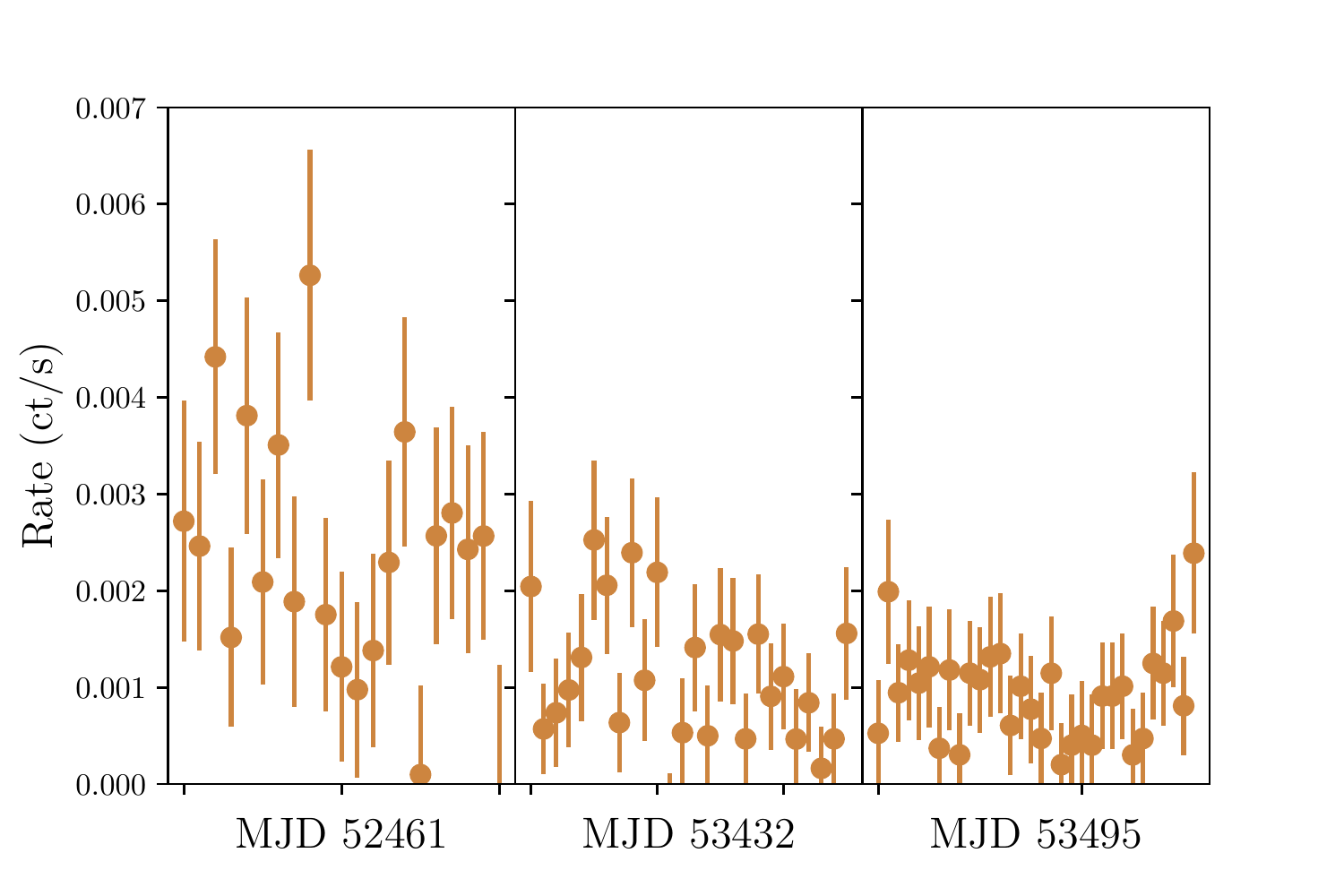} % second figure itself
    \end{minipage}
        \caption{Lightcurves from M87-GCULX3 (right) M87-GCULX4 (left) in ObsIDs 2707, 5826 and 5827, binned by 5ks.}
        \label{fig:firstlcs3}

\end{figure*}

%%%%%%%%%%%%%%%%%%%%%%%%%%%%%%%%%%%%%%%%%%%%%%%%%%



%\begin{thebibliography}{99}
%\end{thebibliography}

%%%%%%%%%%%%%%%%%%%%%%%%%%%%%%%%%%%%%%%%%%%%%%%%%%

%%%%%%%%%%%%%%%%% APPENDICES %%%%%%%%%%%%%%%%%%%%%

\bsp	% typesetting comment
\label{lastpage}
\end{document}